%% file: main.tex
\documentclass[11pt]{article}
\usepackage[letterpaper, margin=1in]{geometry}
\usepackage{amsmath, amssymb, amsthm}
\usepackage{bbm}
\usepackage{graphicx}
\usepackage{url}
\usepackage{thmtools,thm-restate}
\usepackage{hyperref}
\usepackage[capitalise]{cleveref}
\usepackage{algorithm}
\usepackage[noend]{algpseudocode}
\usepackage{subcaption}
\usepackage{xspace}
\usepackage{mathrsfs}
\usepackage{mathtools}
\usepackage{xcolor}
\usepackage{tabularx}
\usepackage{booktabs}
\usepackage{tikz}
\usetikzlibrary{decorations.pathreplacing}
\tikzset{
  exchange vertex/.style={
    circle,draw,fill=white,minimum size=8mm,
    inner sep=1pt,line width=.6pt
  },
  M edge/.style={
    draw=black,solid,line width=.8pt
  },
  sigma edge/.style={
    draw=black,dash pattern=on 4pt off 2.5pt,line width=.8pt
  },
  shortcut edge/.style={
    sigma edge,line width=1.1pt
  },
  row label/.style={
    font=\small\sffamily,anchor=east
  },
  note/.style={
    font=\footnotesize,align=center
  }
}

\input{preamble}

\newtheorem{example}[theorem]{Example}
\newtheorem{property}[theorem]{Property}
\crefname{property}{Property}{Properties}
\Crefname{property}{Property}{Properties}
\usepackage{enumitem}
\usepackage{needspace}
\hypersetup{
  colorlinks=true,
  linkcolor=black,
  filecolor=blue,
  urlcolor=blue,
  citecolor=blue,
  pdfpagemode=FullScreen,
}
\renewcommand{\P}{\mathbb P}
\usepackage[natbib=true, url=false, style=alphabetic, backref=true, doi=false, isbn=false, maxbibnames=9, maxalphanames=4,minalphanames=3]{biblatex}
\renewbibmacro*{pageref}{\iflistundef{pageref}
    {}
    {\setunit{\addperiod\space}\begingroup
     \hypersetup{linkcolor=blue}\printtext{\printlist[pageref][-\value{listtotal}]{pageref}}\endgroup
     \nopunct}}
\AtEveryBibitem{\clearfield{note}
  \clearname{editor}
  \clearlist{language}
  \clearfield{location}
  \clearfield{pages}
}

\IfFileExists{references.bib}{\addbibresource{references.bib}}{}
\newcommand{\Exp}{\mathrm{Exp}}
\providecommand{\OPT}{\mathrm{OPT}}
\providecommand{\ADV}{\mathrm{ADV}}
\providecommand{\ALG}{\mathrm{ALG}}

\providecommand{\E}{\mathbb{E}}

\renewcommand{\vec}[1]{{\mathbf #1}}

\title{The Online $k$-Taxi Problem Is Competitive on All Metrics}
\author{Christian Coester\thanks{Funded by the European Union (ERC, CCOO, 101165139). Views and opinions expressed are however those of the author(s) only and do not necessarily reflect those of the European Union or the European Research Council. Neither the European Union nor the granting authority can be held responsible for them. Email:~\texttt{christian.coester@cs.ox.ac.uk}.}\\University of Oxford\and Yichen Huang\footnote{Supported in part by NSF grant CNS-2107078. Email: \texttt{yichenhuang@g.harvard.edu}.}\\Harvard University\and Tze-Yang Poon\footnote{Email:~\texttt{tpoon@andrew.cmu.edu}.}\\Carnegie Mellon University}
\date{\today}
\begin{document}
\maketitle
\begin{abstract}
    The online $k$-taxi problem is a natural generalization of the $k$-server problem: $k$ taxis must serve a sequence of requests in a metric space. Each request is a pair of points $(r,s)$, and the algorithm serves the request by first moving a taxi to $r$ (to pick up a passenger) and then to $s$ (to drop off the passenger). The cost is the total distance traveled by the taxis without a passenger.

    Previous bounds for general $k$ all depended on the number of points $n$ or the aspect ratio $\Delta$ of the metric space, both of which can be infinite on general metrics, and competitive algorithms were previously known on general metrics only for $k \le 3$. In this work, we give a $9^{k-1}$-competitive randomized algorithm against adaptive online adversaries and, as a corollary, an $81^{k-1}$-competitive deterministic algorithm for the online $k$-taxi problem. 
    
\end{abstract}
\newpage
\pagenumbering{arabic}
\section{Introduction}
In the online $k$-taxi problem~\cite{FRR90}, $k$ taxis are located in a metric space $(V, d)$. Each request is a pair $(r, s) \in V \times V$. The algorithm serves the request by first moving a taxi to $r$ to pick up a passenger and then driving the passenger to $s$. An online algorithm must select a taxi to serve the current request without knowledge of future requests. The classical $k$-server problem~\cite{MMS90} is the special case of $k$-taxi in which $r=s$ for every request.

There are two versions of the problem, depending on which travel distances are included in the cost. In the ``easy'' version, the cost is simply the total distance traveled by the taxis, and the problem has exactly the same competitive ratio as the $k$-server problem~\cite{CK19}. In the ``hard'' version of the problem, the cost is the total distance traveled to $r$, and the distance from $r$ to $s$ is excluded. The motivation is that the distance from $r$ to $s$ is traveled by all algorithms regardless of their decisions, so the hard version charges only the overhead cost. Note that while the (offline) optimal strategy is unchanged, the smaller costs lead to a larger competitive ratio. Hereafter, we mean the hard version when we refer to the $k$-taxi problem.

Despite the interest in the problem, a competitive ratio depending only on $k$ was previously known on general metrics only for $k\le 3$~\cite{Kosoresow96,CK19,CP26}. Other results give bounds that depend on parameters of the metric space. On an $n$-point metric with aspect ratio at most $\Delta$, the randomized algorithm of~\cite{CK19} uses HST embedding techniques and is $O(2^k \log n)$-competitive against oblivious adversaries. \cite{BBCS21} obtains an $O(n^2\log^2 k\log n)$-competitive randomized algorithm through the framework of metrical service systems with transformations. The algorithm of~\cite{BCN23} achieves a competitive ratio of $2^{O(\sqrt{\log k \log \Delta})}\log_{\Delta} n$. \cite{GKP24} gives a randomized $O(\log^3\Delta\log^2(nk\Delta))$-competitive algorithm, using a covering relaxation on HSTs and a primal-dual algorithm. \cite{CH26} uses online embedding techniques to achieve a competitive ratio of $O(2^k \log^2 m)$, where $m \le n$ is the number of distinct points appearing in the initial configuration or the request sequence. Note that all competitive ratios involving $n$ or $\Delta$ are vacuous even on simple metrics like the line or interval [0,1], where $n$ and $\Delta$ are infinite.

\subsection{Our Results}

In this work, we present the first algorithm, more than 35 years after the introduction of the problem~\cite{FRR90}, for online $k$-taxi on general metrics whose competitive ratio depends only on $k$.

\begin{theorem}\label{thm:main}
On every metric space, there is a randomized algorithm for the online $k$-taxi problem with competitive ratio $9^{k-1}$ against adaptive online adversaries.
\end{theorem}

The ratio $9^{k-1}$ is exactly tight for $k\le2$~\cite{CK19}.\footnote{\cite{CK19} proved a lower bound of $9$ for deterministic $2$-taxi algorithms, and we extend it to randomized algorithms against adaptive online adversaries in Appendix~\ref{app:k=2}.} Together with the $2^k-1$ lower bound for randomized algorithms against adaptive online adversaries~\cite{CK19}, our result establishes that the optimal competitive ratio is $2^{\Theta(k)}$ and is tight up to a constant factor in the exponent. Because our proof employs an explicit potential function, we can derandomize the algorithm constructively~\cite{BBKTW94}.

\begin{corollary}\label{cor:det}
    On every metric space, there is a deterministic algorithm for the online $k$-taxi problem with competitive ratio $81^{k-1}$.
\end{corollary}

Our algorithm behind~\cref{thm:main} draws inspiration from the Harmonic algorithm for the $k$-server problem~\cite{Grove91,BG00}. The Harmonic algorithm serves a request at $r$ with a server chosen with probability inversely proportional to its distance from $r$. Instead, our algorithm additionally maintains a ranking among the taxis and assigns weight $w_i$ to the taxi with rank $i$, defined by $w_1=1$ and $w_i=2\cdot3^{i-2}$ for $2\le i\le k$. For a request $(r,s)$, we select the serving taxi with probability proportional to its weight divided by its distance to $r$. The ranking is updated after each request so that, for each $1 \le m \le k$, the locations of the taxis with ranks $1,\dots,m$ are precisely the locations occupied by the same algorithm for the $m$-taxi problem.\footnote{An analogous property is known to be satisfied, for example, by Belady's \textsc{FiF} (farthest-in-future) algorithm for offline paging~\cite{Belady66}: Denoting by $\textsc{FiF}_k^t$ the set of pages in the cache of that algorithm at time $t$ when run with cache size $k$, \cite{BCKPV22} observed that $\textsc{FiF}_k^t\subset \textsc{FiF}_{k+1}^t$ for all $k$ and $t$ for any request sequence (provided the containment holds initially for $t=0$). While the property is a consequence of the simple greedy description of \textsc{FiF}, we use this property as a motivating design principle for our algorithm.} In particular, the taxi that serves the request moves to rank $1$, where its new weight makes it less likely to serve the next request. Other taxis' ranks (and thus weights) can only increase while they are not serving requests. Thus, taxis tend to increase their likelihood of serving requests the longer they remain idle. 

The update rule of the ranking allows us to analyze the change in each prefix of the $m$ smallest ranks separately. As a result, we can prove competitiveness using a potential equal to a weighted sum of the minimum injective matching costs from each prefix of $m$ taxis to the $k$ offline taxis. This gives a new view of potentials involving online and offline configurations: partial configurations of different sizes become building blocks for analyzing the full configuration. Our ranking makes these partial configurations evolve as consistent smaller instances of the same algorithm, creating a hierarchy that enables the analysis. We believe this technique is of independent interest more generally for designing competitive online algorithms.

Note that while the Harmonic algorithm is memoryless, our algorithm uses the ranking of the taxis as additional memory. In fact, using memory is necessary, as we show that no lazy memoryless\footnote{See \cref{sec:memoryless} for detailed definitions of memorylessness and laziness. Note that without laziness, algorithms on continuous metrics may encode their history in the lower-order bits of the taxis' configuration.} algorithm can be competitive even for $k=2$ on the line. This may be surprising because memory is not needed to obtain competitive algorithms for the $k$-server problem on general metrics due to the Harmonic algorithm~\cite{Grove91,BG00}, nor for the $k$-taxi problem on ultrametrics/HSTs~\cite{CK19}.

\begin{restatable}{theorem}{memorylesslowerbound}
\label{thm:memoryless-line-lower-bound}
    On the real line, every lazy memoryless randomized algorithm for the online $2$-taxi problem has infinite competitive ratio, even against an oblivious adversary.
\end{restatable}

\subsection{Related Work}
\label{sec:related-work}
The $k$-server problem, corresponding to the special case in which $r=s$ for each request, is one of the most prominent problems in online algorithms. Deterministically, the competitive ratio for $k$-server is exactly $k$ \cite{MMS90, KP95, CKZ26}. With randomization against oblivious adversaries, polylog$(n)$-competitive algorithms were obtained in~\cite{BBMN15,BCLLM18}, and OpenAI~\cite{OAI26} recently announced one can achieve $O(\log^2 k)$, matching the lower bound of $\Omega(\log^2 k)$ in~\cite{BCR23}. The reader is referred to \cite{Kou09} for a survey on the $k$-server problem. Besides $k$-taxi, many variants of the $k$-server problem have been studied, such as
the $(h,k)$-server problem~\cite{BEJK17,BBCJ20},
the infinite server problem~\cite{CKL17},
the weighted and generalized $k$-server problem~\cite{FR94,BEK17,BMC26,KT04,SS06,BEKN18}, and online multiserver convex chasing and clustering~\cite{BRS21,CHM27}.

All of the aforementioned $k$-server variants (not including $k$-taxi) belong to the general class of metrical task systems (MTS)~\cite{BLS92}, whose tight worst-case competitive ratios on $n$-point metric spaces are $\Theta(n)$ for deterministic algorithms and $\Theta(\log^2 n)$ for randomized algorithms~\cite{BCLL21,CL22MTS,BCR23}. The free passenger relocations in $k$-taxi are naturally captured by the extension to metrical service systems with transformations introduced by~\cite{BBCS21}.

In a metrical service system, an algorithm must move to a point in each requested set. The version with sets of cardinality at most $k$~\cite{CL91Games} is equivalent to width-$k$ layered graph
traversal~\cite{PY91,FFKRRV98}. The reduction in~\cite{CK19} from layered graph traversal to $k$-taxi
establishes a connection that yields exponential deterministic lower
bounds for the latter. This deterministic lower bound is further sharpened in~\cite{CT26}.

\section{Preliminaries}
\label{sec:prelim}
Let $M=(V,d)$ be a metric space. A request is an ordered pair $(r,s)\in V\times V$ and requires a taxi to be moved to $r$ (i.e., picking up the passenger) and then relocated to $s$ (i.e., transporting the passenger to $s$). Driving to $r$ costs the distance traveled; carrying the passenger from $r$ to $s$ is free.

We measure performance using the competitive ratio against an adversary who generates the requests and also serves them. An \emph{oblivious adversary} must fix the request sequence ahead of time (without observing the outcome of the algorithm's random choices) and serves the sequence optimally. An \emph{adaptive online adversary} can choose each request based on the algorithm's past actions, but must serve it immediately (and thus may not be optimal in hindsight). Note that the adaptive online adversary is more powerful than the oblivious adversary. Since our upper bound holds against adaptive online adversaries and the lower bound for memoryless algorithms against oblivious adversaries, both bounds therefore also extend to the other adversary model. For deterministic algorithms, the two adversary models are equivalent. We use $\ADV$ to denote the adversary and $\ALG$ to denote our algorithm. With a slight abuse of notation, we also use these symbols to denote their respective total costs. We write $\ADV_t$ and $\ALG_t$ for the respective costs incurred by the adversary and the algorithm in round $t$.

A randomized algorithm is $\rho$-competitive against an adversary $\ADV$ if for the sequence generated by the adversary, $\E[\ALG] \le \rho \cdot \E[\ADV] + \beta$, where $\beta$ is a constant independent of the request sequence. For deterministic algorithms, the expectations can be dropped and $\ADV$ replaced by an optimal offline algorithm.

For a multiset $X$ and an element $x$, we write $X + x\coloneqq X \cup \set{x}$ and $X - x\coloneqq X \setminus \set{x}$ to insert or remove a single copy of $x$, respectively.

We will use the exponential distribution. A random variable $\zeta$ is sampled from an exponential distribution with rate $\lambda>0$, written $\zeta \sim \Exp(\lambda)$, if the probability density function of $\zeta$ is
\[
f(\zeta) = \begin{cases} \lambda e^{-\lambda \zeta} & \zeta \ge 0,\\
0 & \zeta < 0.
\end{cases}
\]

We will use the following standard fact about exponential races (see, e.g., \cite[Equation 2.10] {durrett2012essentials} and \cite[Lemma 8.5]{mitzenmacher2017probability}): among independent exponential random variables, the probability that a given variable is the smallest is proportional to its rate.
\begin{theorem}
\label{thm:race}
    Let $\zeta_1, \ldots, \zeta_m$ be independent random variables with $\zeta_i \sim \Exp(\lambda_i)$ and $\lambda_i>0$. Then,
    \[
    \P\left(\zeta_i = \min_{j=1}^m \zeta_j \right) = \frac{\lambda_i}{\sum_{j=1}^m \lambda_j}.
    \]
\end{theorem}

Throughout this paper, we assume $k \ge 2$, since $k=1$ is trivial.

\section{Algorithm Description}
We maintain an initially arbitrary ordered list of the taxis. On receiving a request, we update the ordered list and serve the request with the first taxi in the new ordered list. We use $\vec x = (x_1, \ldots, x_k)$ to denote the configuration, which is an ordered list of locations of the taxis. The $i$th taxi in the ordered list has weight $w_i$, defined by $w_1=1$ and $w_i=2\cdot3^{i-2}$ for $2\le i\le k$. These weights are chosen to satisfy $w_{i}=2\sum_{j=1}^{i-1}w_j$ for $2\le i\le k$.

On receiving the request $(r, s)$, let $d_i\coloneqq d(x_i,r)$. For each taxi $x_i$ such that $d_i>0$, draw an
independent exponential random variable $\zeta_i$ of rate $w_i/d_i$. The value of $\zeta_i$ for a taxi $x_i$ already at $r$ is defined to be zero. 
Over the course of the ordered list update, we will swap the positions of the taxis in the ordered list, but these $\zeta_i$ values remain attached to the corresponding \emph{taxis} until the update is completed.

Update the ordered list by scanning the taxis in their original order $i=2,\ldots,k$.
If taxi $x_i$ has a smaller $\zeta_i$ value than the taxi currently in the first position, swap the positions of
these two taxis in the ordered list; otherwise, do nothing. Finally, serve the request with the taxi in the first position. Pseudocode for the algorithm appears in \cref{alg:k-taxi}.

\begin{algorithm}
\caption{Online $k$-Taxi}\label{alg:k-taxi}
\begin{algorithmic}
\Require $k \geq 1$, $w_1=1$, and $w_i=2\cdot3^{i-2}$ for $2\le i\le k$
\State Maintain an initially arbitrary ordered list $(x_1, \ldots, x_k)$.
\While {receiving requests $(r, s)$}
\For {$i=1,\ldots,k$}
\If{$d(x_i,r) = 0$}
\State $\zeta_i \gets 0$
\Else
\State Independently sample $\zeta_i \sim \Exp(w_i/d(x_i, r))$
\EndIf
\If {$\zeta_i < \zeta_1$}
\State Swap $(x_1, x_i)$ and $(\zeta_1, \zeta_i)$.
\EndIf
\EndFor
\State Serve the request with $x_1$.
\EndWhile
\end{algorithmic}
\end{algorithm}

For $1\le m\le k$, call the first $m$ positions the \emph{level-$m$ prefix}. Within this prefix of the original ordered list, call the taxi with the smallest $\zeta_i$ value the \emph{local winner}. The taxi with the smallest $\zeta_i$ value overall is called the \emph{global winner}.

The algorithm always serves the request with the global winner. When some $d_i = 0$, it always uses a taxi already at $r$, breaking ties in favor of smaller original ranking. When all $d_i>0$, it selects taxi $i$ as the global winner with probability proportional to $w_i/d_i$ by \cref{thm:race}. Crucially, the same selection rule applies \emph{at every level}, as established in the following claim.

\begin{claim}[Effect at each level]\label{cl:prefix}
For $1\le m\le k$, let $j_m = \arg\min_{1\le i\le m} \zeta_i$ be the index of the local winner in the original level-$m$ prefix, breaking ties by the smallest index. Let $X_m$ be the multiset of taxi locations in this prefix. After the algorithm updates the list and moves a taxi to $r$, the multiset of locations in the new level-$m$ prefix is $X_m - x_{j_m} + r$.
\end{claim}

\begin{proof}
After the scan has processed slots $1,\ldots,m$, these slots still contain exactly the original first $m$ taxis, and taxi $x_{j_m}$ is at the front. Each later swap involves the taxi in the first position. After the global winner moves to $r$, the multiset of locations in these slots is therefore $X_m - x_{j_m} + r$.
\end{proof}

In words, a prefix ending before the global winner's original position loses its local winner and gains the global winner, which is then moved to $r$, as depicted in \cref{fig:permutation-update}. Thus, for any fixed $m$, the multiset of \emph{locations} in the level-$m$ prefix changes as if \emph{its local winner were moved to $r$}. This viewpoint will be helpful for understanding the analysis. Note that here we refer to the locations of the taxis in the level-$m$ prefix, instead of the identity of the taxis.

\begin{example}
    We give a simple example to illustrate the update rule and the prefix property. Suppose the current ordered configuration is $(x_1,x_2,x_3,x_4)$ and the corresponding $\zeta$ values are $(8,3,5,1)$. The update produces $(x_4,x_1,x_3,x_2)$: it first swaps $x_2$ with $x_1$, leaves $x_3$ in place, and then swaps $x_4$ with $x_2$.

    The algorithm then moves $x_4$ to $r$ for the pickup. The multisets of locations in the level-$1$, level-$2$, and level-$3$ prefixes become $\set{r}$, $\set{r,x_1}$, and $\set{r,x_1,x_3}$, respectively. At levels $2$ and $3$, these multisets change exactly as if the local winner $x_2$ had performed the pickup. The actual serving taxi is $x_4$, and the ordered configuration after the pickup is $(r,x_1,x_3,x_2)$.
\end{example}

A consequence of~\cref{cl:prefix} is that the locations of the first $m$ taxis in the list have exactly the same distribution as those of the algorithm with $m$ taxis running on the same request sequence: one can couple the evolution of the first $m$ taxi locations with the $m$-taxi run.

Note that all taxis except the global winner will stay in place or move toward the end of the ordered list. Thus, while we place the global winner first and serve the request with it, a taxi's position in the ordered list is in general not monotone with respect to its $\zeta_i$ value. On the other hand, being placed later in the ordered list increases a taxi's weight in future selections, giving more priority to a taxi that ``almost'' serves.

\input{figures/permutation-update}

\section{The Analysis}
\label{sec:analysis}

The central quantity in our analysis is the minimum injective matching cost. For finite multisets $X, Y$ of locations in $V$ with $|X| \le |Y|$, define
\[
D(X,Y)\coloneqq\min_{\sigma\colon X\hookrightarrow Y}\sum_{x\in X}d(x,\sigma(x))
\]
to be the minimum injective matching cost from $X$ to $Y$. For an online configuration $\vec x$, which is an ordered sequence of taxi locations, and an adversary configuration $Y$, we consider a potential of the form
\[\Phi(\vec x,Y)=\sum_{m=1}^{k}\alpha_m D(X_m, Y),\]
where $X_m \coloneqq \{x_1, \ldots, x_m\}$ is the multiset of taxi locations in the level-$m$ prefix, and $\alpha_1>\dots>\alpha_k>0$ are coefficients that will be chosen later.

\subsection{Setup}

Consider a round $t$ where a request $(r,s)$ arrives, and write $\ALG_t$ and $\ADV_t$ for the respective movement costs of our algorithm and the adversary. We analyze round $t$ in three steps: First, the adversary moves a taxi to $r$. Next, the algorithm updates its ordered list and moves a taxi to $r$. Finally, both algorithms move their respective taxis from $r$ to $s$.

We choose the coefficients $\alpha_m$ to satisfy the following property: in the second step, where the adversary already has a taxi at $r$, the potential drops enough to pay for the algorithm's movement.

\begin{property}
\label{prop:online-potential}
    For every round $t$ with request $(r,s)$, every online configuration, and every adversary configuration containing $r$,
    \[
    \E[\Delta \Phi + \ALG_t] \le 0,
    \]
    where $\Delta\Phi$ is the potential change during the algorithm's ordered list update and movement to $r$.
\end{property}

For appropriately chosen $\{\alpha_m\}$ such that \cref{prop:online-potential} holds, \cref{alg:k-taxi} is competitive:

\begin{lemma}
\label{lem:implication}
    If \cref{prop:online-potential} holds, then \cref{alg:k-taxi} is $(\sum_m \alpha_m)$-competitive against adaptive online adversaries.
\end{lemma}
\begin{proof}
    Fix a round $t$ with request $(r,s)$ and condition on the history and the adversary's current move. Analyze the round in the steps discussed above, writing $\ALG_t$ and $\ADV_t$ for the respective movement costs.
    
    \begin{itemize}[noitemsep,nosep]
        \item First, the adversary moves a taxi to $r$. Since matching costs are 1-Lipschitz, $\Delta \Phi \le \sum_m \alpha_m \ADV_t$. 
        \item Next, the algorithm moves a taxi to $r$. \Cref{prop:online-potential} ensures that $\E[\Delta\Phi + \ALG_t] \le 0$.
        \item Finally, both move their respective taxis from $r$ to $s$. We have $\Delta \Phi = 0$, since an optimal injection can be chosen to match the online and offline taxis that serve the request to each other both before and after the relocation.
    \end{itemize}

    Taking full expectations and combining the three steps gives, in every round $t$, \[\E[\Delta \Phi + \ALG_t] \le \Bigl(\sum_m \alpha_m\Bigr)\E[\ADV_t].\] Summing over all rounds and noting that the initial potential is $0$ and the final potential is nonnegative gives $\E[\ALG] \le (\sum_m \alpha_m)\E[\ADV]$.
\end{proof}

It thus remains to prove \cref{prop:online-potential}. We begin by bounding the online movement cost. Fix the pickup location $r$ and $1 \le m \le k$. Recall that $d_i = d(x_i, r)$ for $1 \le i \le k$. When $d_1,\ldots,d_m>0$, let $Z_m\coloneqq \sum_{i=1}^m w_i/d_i$ be the normalizing constant for level $m$.

\begin{lemma}
\label{lem:online-move}
    Fix a round $t$ with request $(r, s)$ and condition on the current configuration. If $d_i > 0$ for all $1 \le i \le k$, then $\E[\ALG_t] = 3^{k-1}/Z_k$. Otherwise, if $d_i = 0$ for some $i$, $\ALG_t = 0$.
\end{lemma}
\begin{proof}
    If $d_i = 0$ for some $i$, then $\zeta_i = 0$ and the global winner must already be at $r$.     Otherwise, by \cref{thm:race},
    \[
    \E[\ALG_t] = \sum_{i=1}^k \P(j_k = i) \cdot d_i = \sum_{i=1}^k \frac{w_i}{d_i Z_k} d_i = \frac{1}{Z_k} \sum_{i=1}^k w_i = \frac{3^{k-1}}{Z_k}. \qedhere
    \]
\end{proof}

\subsection{Proof Idea}
Fix a round $t$ with request $(r,s)$, an online ordered list, and an adversary configuration $Y$ containing $r$. Recall that $X_m$ is the multiset of taxi locations in the online algorithm's level-$m$ prefix. As established in \cref{cl:prefix}, we can reason about each level $m$ separately. We bound $\E[\Delta D(X_m,Y)]$, the expected change in the level-$m$ matching cost during the online algorithm's ordered list update and movement to $r$. We first assume that $d_i > 0$ for all $1\le i\le k$ and will handle zero distances later.
We say a matching \emph{uses} $r$ if it matches some $x_i \in X_m$ to $r \in Y$, and \emph{omits} $r$ otherwise.

For $1\le i\le m$, let $\Delta_{m,i}$ be the change in $D(X_m,Y)$ if taxi $x_i$ were moved to $r$:
\[
\Delta_{m,i}\coloneqq D(X_m - {x_i} + r,Y) - D(X_m,Y) = D(X_m-x_i, Y-r)-D(X_m, Y).
\]
By~\cref{cl:prefix} and the exponential race (\cref{thm:race}), the local winner of the level-$m$ prefix is $x_i$ with probability $w_i/(d_iZ_m)$, in which case $X_m$ becomes $X_m-x_i+r$. Thus, the expected change in the potential is
\[
\E[\Delta\Phi]=\sum_{m=1}^k \frac{\alpha_m}{Z_m} \sum_{i=1}^m \frac{w_i}{d_i} \Delta_{m, i}.
\]

We fix an optimal matching $M_m$ from $X_m$ to $Y$ at each level $m$ just after the offline algorithm's movement by building it successively. Start from $M_0=\varnothing$. At level $m$, augment $M_{m-1}$ along an augmenting path from $x_m$ to an unused offline vertex $y_m$ that minimizes the increase in matching cost, obtaining an optimal matching $M_m\colon X_m\hookrightarrow Y$ with image $Y_m=\{y_1,\dots,y_m\}$. The images $Y_m$ form a nested sequence. Here and onwards, in slight abuse of notation we use $M_m$ interchangeably for the function from $X_m$ to $Y$ and for the set of edges of the corresponding matching. 
Note that after the online update, the optimal target multisets $Y_m$ may differ, so we will consider injections directly into $Y$ after the update. 
Let $\ell\in[k]$ be the first index with $y_\ell=r$. In other words, levels $1,\dots,\ell-1$ omit $r$, and levels $\ell,\dots,k$ use $r$.

\paragraph{Analysis with a simplifying assumption.}

To illustrate the main idea of the proof, we first give a sketch with the additional assumption that $x_\ell$ is matched to $r$ in the chosen optimum matching $M_\ell$. Since $\ell$ is the first level whose matching uses $r$, one might naively think that $x_\ell$ is matched to $r$.

For levels $m<\ell$, the optimum matching from $X_m$ to $Y$ omits $r$. Hence, if any taxi in $X_m$ is moved to $r$, it can subsequently be matched to $r$ at cost $0$ and the cost of the level-$m$ matching does not increase, i.e., $\Delta_{m,i}\leq 0$ for all $i$. This allows us to upper bound the total expected change in the potential from levels $m<\ell$ by
\begin{align*}
    \sum_{m=1}^{\ell-1}\frac{\alpha_m}{Z_m}\sum_{i=1}^m\frac{w_i}{d_i}\Delta_{m,i}
    \leq 0.
\end{align*}

For levels $m > \ell$, by the triangle inequality, $D(X_m-x_i+r, Y) \leq D(X_m,Y) + d_i$, so $\Delta_{m,i}/d_i\leq 1$.
Consequently, we can upper bound the total expected change in potential from levels $m>\ell$ by
\begin{align*}
    \sum_{m=\ell+1}^k\frac{\alpha_m}{Z_m}\sum_{i=1}^m\frac{w_i}{d_i}\Delta_{m,i}
    \leq \sum_{m=\ell+1}^k\frac{\alpha_m}{Z_m}\sum_{i=1}^m w_i
    = \sum_{m=\ell+1}^k\frac{3^{m-1}}{Z_m}\alpha_m.
\end{align*}

At level $\ell$, under the assumption that $x_\ell$ is matched to $r$, the movement of $x_\ell$ to $r$ decreases the cost of the level-$\ell$ matching by exactly $d_\ell$, i.e., $\Delta_{\ell,\ell}=-d_\ell$. The expected change in potential at level $\ell$ is therefore at most
\begin{align*}
    \frac{\alpha_\ell}{Z_\ell} \sum_{i=1}^\ell\frac{w_i}{d_i}\Delta_{\ell,i}
    \leq \frac{\alpha_\ell}{Z_\ell} \sum_{i=1}^{\ell-1}w_i - \frac{\alpha_\ell w_\ell}{Z_\ell}
    \leq -\frac{3^{\ell-2}}{Z_\ell}\alpha_\ell,
\end{align*}
where we use ${\Delta_{\ell,i}}/{d_i}\leq 1$ and $\sum_{i=1}^{\ell-1}w_i\leq w_{\ell}/2$.

Recall from \cref{lem:online-move} that the expected online movement cost is $3^{k-1}/Z_k$.
Combining all bounds above and using $Z_j\geq Z_\ell$ for $j\geq\ell$ gives
\[
\E[\Delta\Phi + \ALG_t] \leq \sum_{m=\ell+1}^k\frac{3^{m-1}}{Z_m}\alpha_m -\frac{3^{\ell-2}}{Z_\ell}\alpha_\ell + \frac{3^{k-1}}{Z_k} \leq \frac{1}{Z_\ell}\left(\sum_{m=\ell+1}^k 3^{m-1}\alpha_m -3^{\ell-2}\alpha_\ell + 3^{k-1}\right).
\]

Choosing the coefficients $\alpha_m$ to decrease geometrically and sufficiently rapidly with $m$ makes the final expression nonpositive and establishes \cref{prop:online-potential}, completing the proof sketch of competitiveness (albeit slightly weaker) under the simplifying assumption.

\paragraph{Roadmap for the general analysis.} The main challenge is that the assumption that $x_\ell$ is matched to $r$ at level $\ell$ is not true in general, so the decrease in potential due to level $\ell$ alone may not suffice. However, in the sketch above, we only used the nonpositivity of potential changes for levels $m<\ell$. 
In the general analysis, we will keep the decrease in potential at these levels, and show that a weighted sum of these earlier decreases bounds how far the matching-cost decrease at level $\ell$ falls short of $d_\ell$.

To see the high-level idea, note that the symmetric difference $M_{\ell-1}\triangle M_\ell$ is an alternating path from $r$ to $x_\ell$. If the edges of $M_{\ell-1}$ on this path are short, alternating $M_\ell$ along this path would give a decrease close to $d_\ell$, even if $x_\ell$ was not directly matched to $r$. In general, the gap to the desired decrease $d_\ell$ is at most twice the total length of the edges of $M_{\ell-1}$ on this path. On the other hand, if the edges of $M_{\ell-1}$ have large lengths, we can show that the decreases in matching costs of levels $m<\ell$ are large.

Removing the assumption in this way would give a competitive ratio of $O(12^k)$. To obtain the sharper bound of $9^{k-1}$, we apply the same argument at every level $m\ge\ell$, rather than only at level $\ell$. For each such $m$, we can find a taxi $x_j$ with $\ell\le j\le m$ whose movement decreases the matching cost at level $m$ by nearly $d_j$ when certain edges of $M_{\ell-1}$ are short. This allows us to choose coefficients $\alpha_m$ whose sum is $9^{k-1}$, as promised.

\subsection{Actual Potential Changes at Each Level}\label{sec:level-changes}

\input{figures/exchange-optimal}
For each level $m\ge\ell$, we seek a taxi $x_j$, with $\ell\le j\le m$, whose movement to $r$ decreases the matching cost by nearly $d_j$. Equivalently, we want the normalized decrease $-\Delta_{m,j}/d_j$ to be close to $1$. This taxi will be the endpoint of the alternating path from $r$ in $M_m\cup M_{\ell-1}$. Alternating $M_m$ along this path reduces the matching cost by the length of the path minus twice the length of the $M_{\ell-1}$-edges on this path.

Thus, we want to upper bound the length of the $M_{\ell-1}$-edges on such an alternating path, relative to an unknown $d_j$. Let
\[
\varepsilon_{\ell-1}\coloneqq
\min\left\{\varepsilon\ge0\colon
  \sum_{\substack{i\le \ell-1\\d_i\le T}}d(x_i,M_{\ell-1}(x_i))
  \le\varepsilon T
  \ \text{for every }T>0
\right\}.
\]
As we show later (in \Cref{cl:tight-late-drift}), a small value of $\varepsilon_{\ell-1}$ implies that the edges of $M_{\ell-1}$ account for only a small fraction of the total length of any such alternating path from $r$. See~\cref{fig:optimal-matching-exchange} for an illustration.

Write
\[
\Delta_m\coloneqq\sum_{i=1}^m\frac{w_i}{d_i}\Delta_{m,i}
\qquad(1\le m\le k),
\]
so that $\Delta_m/Z_m$ is the expected change in the level-$m$ matching cost. We will show:
\begin{itemize}
    \item (\Cref{cl:tight-late-drift}) At each level $m\ge\ell$, there is a taxi of rank $\ge \ell$ whose movement to $r$ decreases the matching cost by at least $(1-2\varepsilon_{\ell-1})$ times its distance to $r$. Bounding the changes from the other taxis using $\Delta_{m, i} /d_i \le 1$ gives
    \[
    \Delta_m\le3^{m-1}
    -2\bigl(1-\min\{\varepsilon_{\ell-1},1/2\}\bigr)w_\ell.
    \]
    \item (\Cref{cl:epsilon-bound}) A weighted sum of the matching-cost decreases at levels $m<\ell$ bounds the capped value of $\varepsilon_{\ell-1}$:
    \[
    \min\{\varepsilon_{\ell-1},1/2\}
    \le-\sum_{m=1}^{\ell-1}\frac{3^{\ell-m-1}}{w_m}\Delta_m.
    \]
    Thus, when the first bound becomes weaker, these lower-level decreases can compensate. 
\end{itemize}
These bounds make the sketch precise without the assumption. When $\varepsilon_{\ell-1}=0$, the first bound recovers the decrease at level $\ell$ under the assumption and improves the estimate for levels $m>\ell$. Since the expected potential change is $\E[\Delta\Phi]=\sum_m\alpha_m\Delta_m/Z_m$, we can choose exponentially decreasing coefficients $\alpha_m$ so that the earlier decreases compensate for the loss at later levels.
The cap at $1/2$ marks the point where the alternating-path analysis above no longer guarantees a strict decrease for levels $m\geq \ell$ from the selected taxis' move; in this case, the levels $m<\ell$ provide enough decrease to compensate. 

\paragraph{Levels $\boldsymbol{m \ge \ell}$.} The analysis for levels $m\ge \ell$ is straightforward from the definition of $\varepsilon_{\ell-1}$.
\begin{claim}[Levels $m\ge \ell$]\label{cl:tight-late-drift}
For every $\ell\le m\le k$,
\[
\Delta_m
\le3^{m-1}-2\bigl(1-\min\{\varepsilon_{\ell-1},1/2\}\bigr)w_\ell.
\]
\end{claim}
\begin{proof}
Let $P$ be the alternating path from $r$ in $M_{\ell-1}\cup M_m$. Since the image of $M_{\ell-1}$ is contained in that of $M_m$, the path ends at an online taxi $x_j$ with $\ell\le j\le m$. Let $L$ be the total length of $P$ and $L_{\ell-1}$ the total length of its $M_{\ell-1}$-edges.

Switching $M_{\ell-1}$ along $P$ with its final edge removed gives a feasible matching of $X_{\ell-1}$ into $Y$. Optimality of $M_{\ell-1}$ therefore gives $L-2L_{\ell-1}\ge0$. Moving $x_j$ to $r$ and switching the remaining path edges gives
\[
\Delta_{m,j}\le-L+2L_{\ell-1}\le0.
\]
On the other hand, every old online taxi on $P$ has $d_i\le L$, so the definition of $\varepsilon_{\ell-1}$ gives $L_{\ell-1}\le\varepsilon_{\ell-1}L$. Since $L\ge d_j$, combining this with the nonpositive bound above gives
\[
\Delta_{m,j}\le
-\max\{0,1-2\varepsilon_{\ell-1}\}\,d_j.
\]
Bounding the changes from the other taxis by $\Delta_{m,i}\le d_i$ and using $w_j\ge w_\ell$ now gives
\begin{align*}
    \Delta_m = \sum_{i=1}^m \frac{w_i \Delta_{m, i}}{d_i} &\le \sum_{i=1}^{m}w_i - w_j + \frac{w_j \Delta_{m, j}}{d_j} \\&\le 3^{m-1} -\bigl(1+\max\{0,1-2\varepsilon_{\ell-1}\}\bigr)w_j \\&\le3^{m-1}-2\bigl(1-\min\{\varepsilon_{\ell-1},1/2\}\bigr)w_\ell. \qedhere
\end{align*}
\end{proof}

\paragraph{Levels $\boldsymbol{m < \ell}$.} To show the second property, it will help to extend the definition of $\varepsilon_{\ell-1}$ to all $m < \ell$: for $m < \ell$ and the selected optimum matching $M_m$, define
\[
\varepsilon_{m}\coloneqq
\min\left\{\varepsilon\ge0\colon
  \sum_{\substack{i\le m\\d_i\le T}}d(x_i,M_{m}(x_i))
  \le\varepsilon T
  \ \text{for every }T>0
\right\}.
\]
Recall that, since $M_{m-1}$ omits $r$, $D(X_m, Y)-D(X_{m-1}, Y)$ is exactly the matching-cost decrease at level $m$ if $x_m$ is moved to $r$. Define the normalized matching-cost decrease from moving $x_m$ as
\[
g_m\coloneqq\frac{D(X_m,Y)-D(X_{m-1},Y)}{d_m}\ge0
\qquad(1\le m<\ell).
\]
Then $\Delta_{m,m} = -g_md_m$. The next claim relates this decrease to the evolution of $\varepsilon_m$.

\begin{claim}[Recurrence on $\varepsilon_m$]\label{cl:epsilon-evolution}
For every $1\le m<\ell$, $\min\{\varepsilon_m,1/2\}
\le3\min\{\varepsilon_{m-1},1/2\}+g_m$.
\end{claim}
\begin{proof}

\input{figures/epsilon-radius-entry}
If the right-hand side is at least $1/2$, the bound is immediate. Otherwise, $\varepsilon_{m-1}<1/2$. Set $\varepsilon\coloneqq3\varepsilon_{m-1}+g_m<1/2$. We will show that $\varepsilon_m\le\varepsilon$. For each level, write
\[
C_m(T)\coloneqq\sum_{\substack{i\le m,d_i\le T}}d(x_i,M_m(x_i)).
\]
We will show that $C_m(T) \le \varepsilon T$ for all $T>0$.

Let $P$ be the augmenting path that changes $M_{m-1}$ into $M_m$, starting at $x_m$ and ending at the offline vertex $y_m$ that is unused by $M_{m-1}$. Let $L$ be its total length and $L_{m-1}$ the total length of its old edges. Then
\[
L-2L_{m-1}=g_md_m,\qquad
L_{m-1}\le C_{m-1}(d_m+L)
           \le\varepsilon_{m-1}(d_m+L).
\]
Here every online taxi on $P$ is within distance $d_m+L$ of $r$, and the last inequality follows from the definition of $\varepsilon_{m-1}$. Since $3\varepsilon_{m-1}+g_m<1/2$, these inequalities give
\[
\frac{L}{d_m}
\le\frac{g_m+2\varepsilon_{m-1}}{1-2\varepsilon_{m-1}}
<\frac12,
\qquad
L_{m-1}\le\frac32\varepsilon_{m-1}d_m.
\]
The new edges along $P$ have total length $g_md_m+L_{m-1}$.

If $T\ge d_m$, retaining all unchanged edges and counting every new edge gives
\[
C_m(T)
\le C_{m-1}(T)+g_md_m+L_{m-1}
\le\left(\frac52\varepsilon_{m-1}+g_m\right)T
\le\varepsilon T.
\]

Now let $T<d_m$. If $P$ contains no old online taxi within distance $T$ of $r$, then $C_m(T)=C_{m-1}(T)\le\varepsilon T$. Otherwise, let $L_{m-1}^{>T}$ and $L_m^{>T}$ be the total lengths of the $M_{m-1}$-edges and $M_m$-edges of $P$, respectively, whose online endpoints have distance greater than $T$ from $r$.

Traverse $P$ from $x_m$, stopping at the first online taxi $x_i$ with $d_i\le T$ (see \cref{fig:epsilon-radius-entry} for an illustration). Since $d_m>T$, this initial subpath has length at least $d_m-d_i\ge d_m-T$. An offline vertex may already be inside, but every online taxi before $x_i$ is outside. Thus, all new edges on this subpath are counted by $L_m^{>T}$, and all old edges except the last are counted by $L_{m-1}^{>T}$. The last edge is the old assignment edge of $x_i$, whose length is at most $C_{m-1}(T)$. Therefore,
\[
L_{m-1}^{>T}+L_m^{>T}+C_{m-1}(T)\ge d_m-T.
\]

The change in the cost counted inside distance $T$ is the total matching-cost increase minus the change counted outside. Therefore,
\begin{align*}
C_m(T) &=C_{m-1}(T)+g_md_m+L_{m-1}^{>T}-L_m^{>T} \\&\le(g_m-1)d_m+T+2\bigl(L_{m-1}^{>T}+C_{m-1}(T)\bigr)\\
&\le(g_m-1)d_m+T+2C_{m-1}(d_m+L) \le(\varepsilon-1)d_m+T \le\varepsilon T.
\end{align*}
The bound using $C_{m-1}(d_m+L)$ follows because the old edges counted by $L_{m-1}^{>T}$ and $C_{m-1}(T)$ are disjoint, and all their online endpoints are within distance $d_m+L$ of $r$. 
\end{proof}

Iterating this bound lets us bound the capped value of $\varepsilon_m$ by the earlier matching-cost decreases.

\begin{claim}[Levels $m<\ell$]\label{cl:epsilon-bound}
For every $1\le m<\ell$, $\Delta_m\le0$. Moreover,
\[
\min\{\varepsilon_{\ell-1},1/2\}
\le-\sum_{m=1}^{\ell-1}\frac{3^{\ell-m-1}}{w_m}\Delta_m.
\]
\end{claim}
\begin{proof}
For each $m<\ell$, $M_m$ omits $r$, so $\Delta_{m,i}\le0$ for every $i\le m$. Together with $\Delta_{m,m}=-g_md_m$, this gives $\Delta_m\le-w_mg_m\le0$. Iterating \cref{cl:epsilon-evolution} from $\varepsilon_0=0$ therefore yields
\[
\min\{\varepsilon_{\ell-1},1/2\}
\le\sum_{m=1}^{\ell-1}3^{\ell-m-1}g_m
\le-\sum_{m=1}^{\ell-1}\frac{3^{\ell-m-1}}{w_m}\Delta_m.
\qedhere
\]
\end{proof}

\subsection{\texorpdfstring{Choosing the Coefficients $\boldsymbol{\alpha_m}$}{Choosing the Coefficients α}}
Since $w_i=w_{i+1}/3$ for $2\le i<k$, the factor $3^{\ell-m-1}/w_m$ in \cref{cl:epsilon-bound} suggests tail sums of $\alpha_m$ decreasing by a factor of $9$. Here, we first give a simpler, but slightly worse bound of $3\cdot 9^{k-1}$, and then explain how to tighten it to $9^{k-1}$. In this regard, we choose the tail sums $\sum_{m=j}^k\alpha_m=3\cdot9^{k-j}$ for $1\le j\le k$.

\begin{lemma}\label{lem:loose-bound}
    For $k\ge 2$, \cref{prop:online-potential} holds with
    \[
    \alpha_m=24\cdot9^{k-m-1}\quad(1\le m<k),\qquad
    \alpha_k=3.
    \]
    Consequently, \cref{alg:k-taxi} is $3\cdot 9^{k-1}$-competitive against adaptive online adversaries.
\end{lemma}
\begin{proof}
We first establish~\cref{prop:online-potential} that the potential decrease pays for the movement.
First suppose all $d_i>0$. The contribution from levels $m<\ell$ satisfies
\[
\E\left[\sum_{m=1}^{\ell-1}\alpha_m\Delta D(X_m,Y)\right] =\sum_{m=1}^{\ell-1}\frac{\alpha_m}{Z_m}\Delta_m \le\frac{8w_\ell9^{k-\ell}}{Z_\ell} \sum_{m=1}^{\ell-1}\frac{3^{\ell-1-m}}{w_m}\Delta_m \le-\frac{8w_\ell9^{k-\ell}}{Z_\ell}\min\{\varepsilon_{\ell-1},1/2\}.
\]
The first inequality uses $\alpha_mw_m\ge8w_\ell9^{k-\ell}3^{\ell-1-m}$, which follows from $w_\ell\le3^{\ell-m}w_m$ for $m<\ell$. Since $\Delta_m\le0$ and $Z_m\le Z_\ell$, replacing $Z_m$ by $Z_\ell$ gives an upper bound. The final inequality uses \cref{cl:epsilon-bound}.

For $m>\ell$, the bound in \cref{cl:tight-late-drift} is nonnegative, so replacing its denominator $Z_m$ by $Z_\ell$ preserves an upper bound. The coefficient choice and $w_\ell\ge2\cdot3^{\ell-2}$ give
\[
\sum_{m=\ell}^k\alpha_m=3\cdot9^{k-\ell},
\qquad
\sum_{m=\ell}^k\alpha_m(3^{m-1}-2w_\ell)\le-3^{k-1}
\]
so \cref{cl:tight-late-drift} gives
\begin{align*}
\E\left[\sum_{m=\ell}^{k}\alpha_m\Delta D(X_m,Y)\right]
&\le\frac1{Z_\ell}\sum_{m=\ell}^k\alpha_m
    \bigl(3^{m-1}-2(1-\min\{\varepsilon_{\ell-1},1/2\})w_\ell\bigr)\\
&=\frac1{Z_\ell}\left(
    \sum_{m=\ell}^k\alpha_m(3^{m-1}-2w_\ell)
    +2w_\ell\min\{\varepsilon_{\ell-1},1/2\}\sum_{m=\ell}^k\alpha_m
    \right)\\
&\le\frac{-3^{k-1}
    +6w_\ell9^{k-\ell}\min\{\varepsilon_{\ell-1},1/2\}}{Z_\ell}.
\end{align*}
Combining these bounds with \cref{lem:online-move} yields
\[
\E[\Delta\Phi+\ALG_t]
\le \frac{3^{k-1}}{Z_k} - \frac{3^{k-1}}{Z_\ell} -\frac{2w_\ell9^{k-\ell}}{Z_\ell}\min\{\varepsilon_{\ell-1},1/2\}
\le0.
\]

Now suppose a taxi is already at $r$. Let $x_j$ be the smallest ranked taxi at $r$. Every prefix of level $m \ge j$ is unchanged. If $j \le \ell$, then the expected change in potential at each changing level is nonpositive. Otherwise, levels $\set{1, \ldots, j-1}$ form a sequence where $d_i > 0$. Applying the calculation until $j-1$ gives
\[
\E\left[\sum_{m=1}^{j-1}\alpha_m \Delta D(X_m,Y)\right] \le-\frac{3^{j-2}}{Z_\ell} \le 0.
\]
Here the potential need only be nonincreasing, since the actual movement cost is zero. This completes the proof of~\cref{prop:online-potential}.

Finally, $\sum_m\alpha_m=3\cdot9^{k-1}$, so \cref{lem:implication} gives the competitive bound.
\end{proof}

The ratio $3\cdot 9^{k-1}$ already suffices to certify a competitive algorithm on general metrics. To obtain the sharper bound, we keep the coefficients for $m\ge2$ and set $\alpha_1=2\sum_{m=2}^k\alpha_m$. This requires a separate treatment of the level-$1$ prefix.

\begin{lemma}\label{lem:sharp-bound}
For $k\ge 2$, \cref{prop:online-potential} holds with
\[
\alpha_1=6\cdot9^{k-2},\qquad
\alpha_m=24\cdot9^{k-m-1}\quad(2\le m<k),\qquad
\alpha_k=3.
\]
Consequently, \cref{alg:k-taxi} is $9^{k-1}$-competitive against adaptive online adversaries.
\end{lemma}
\begin{proof}
Only $\alpha_1$ changes from \cref{lem:loose-bound}. If $x_1\in Y$, the level-$1$ matching cost is zero before and after the online update, so reducing $\alpha_1$ leaves the preceding bound unchanged.

Now suppose $x_1\notin Y$, so $d_1>0$. We bound the contribution from levels $m\ge2$ together with the movement cost. Consider the matching partner $y_1$ of $x_1$. 

If $y_1\ne r$, then consider a modified adversary configuration where the location of $y_1$ is changed to $x_1$ for comparison. This preserves the adversary taxi at $r$, so the preceding argument applies to the modified configuration. For all levels $m \ge 2$, this modification affects each matching cost by at most $d(x_1,y_1)$ both before and after the online update. At level $1$, the modified matching cost is zero before and after the online update, whereas the true matching cost decreases from $d(x_1,y_1)$ to zero during the online update. Hence
\[
\E[\Delta\Phi+\ALG_t]
\le\left(2\sum_{m=2}^k\alpha_m-\alpha_1\right)d(x_1,y_1)=0.
\]

It remains to consider the case $y_1=r$, meaning $\ell=1$. Here $\varepsilon_0=0$. If all $d_i > 0$, then using~\cref{cl:tight-late-drift} gives
\[
\E[\Delta\Phi] \le \sum_{m=1}^k\frac{\alpha_m(3^{m-1}-2)}{Z_m} \le -\frac{3^{k-1}}{Z_1} \le -\frac{3^{k-1}}{Z_k},
\]
which is enough to pay the movement cost. Otherwise, let $j$ be the smallest index with $d_j = 0$, then $j \ge 2$ and only levels $m < j$ change. Truncating the same calculation at $j-1$ bounds the change in potential by $-3^{j-2}/Z_{1}$, whereas the movement cost is $0$. This proves \cref{prop:online-potential}.

Finally, $\sum_m\alpha_m=9^{k-1}$, so \cref{lem:implication} gives the competitive bound and proves \cref{thm:main}.
\end{proof}

The ratio $9^{k-1}$ is tight for our algorithm, as shown in Appendix~\ref{sec:lb-algorithm}.

\section{Derandomizing the Algorithm}
\label{sec:derandomization}
The derandomization follows from \cite[Thm 3.1]{BBKTW94}. We provide a sketch here for completeness. Roughly speaking, one can use the distribution of our randomized algorithm as the adversary's distribution.

\begin{proof}[Proof of \cref{cor:det}]
We derandomize by using a virtual copy of our randomized algorithm as the adversary. More precisely,
maintain an ordered configuration $\vec x$ and the exact distribution $\mc Y$
of the taxi locations of an independent virtual copy of \cref{alg:k-taxi}.

On receiving $(r,s)$, first advance the distribution of $\mc Y$ through
its ordered list update and movement, obtaining $\mc Y'$. Among all possible
outcomes $\vec x'$ of applying our randomized update and movement to $\vec x$,
choose one minimizing
$c(\vec x,\vec x')+\E_{Y'\sim \mc Y'}[\Phi(\vec x',Y')]$, where $c(\vec x,\vec x')$ is the movement cost.
Since every configuration $Y'$ contains $r$, \cref{prop:online-potential}, averaged
over $Y'$, guarantees that the chosen outcome satisfies
\[
    c(\vec x,\vec x')+
    \E\!\left[\Phi(\vec x',Y')-\Phi(\vec x,Y')\right]\le 0.
\]
Finally, both the deterministic algorithm and the virtual copy carry the passenger to $s$.

The virtual movement increases $\E_{Y \sim \mc Y}[\Phi(\vec x,Y)]$ by at most $9^{k-1}$ times
its expected cost, whereas the simultaneous passenger relocation
leaves the potential unchanged. Let $\ALG^{\mathrm{det}}$ and $\ALG^{\mathrm{virt}}$ denote the respective total costs of the deterministic algorithm and the virtual copy, and let $\OPT$ be the optimal offline cost for the request sequence. Summing the preceding inequality
therefore gives
\[
    \ALG^{\mathrm{det}}
    \le 9^{k-1}\,\E[\ALG^{\mathrm{virt}}]
    \le 81^{k-1}\cdot \OPT. \qedhere
\]
\end{proof}

\section{Lazy Memoryless Algorithms Are Not Competitive}
\label{sec:memoryless}
In this section, we prove that lazy memoryless algorithms cannot have a bounded competitive ratio. An algorithm is \emph{lazy} if it only moves a single taxi to serve each request $(r,s)$ if there is no taxi at $r$, and uses a taxi at $r$ otherwise. A lazy algorithm is \emph{memoryless} if its distribution over serving locations, conditioned on the history, depends only on the unordered multiset of current taxi locations and the current request.

\memorylesslowerbound*

In the main text, we give a short proof under the additional assumption of reflection
invariance; the full proof without this assumption appears
in Appendix~\ref{app:memoryless}. We call a memoryless algorithm \emph{reflection-invariant} if reflecting all locations about any point $a\in\mathbb R$ preserves the serving probabilities. Explicitly, if it serves $(r,s)$ from $x$ with probability $p$ when the taxis are at $x$ and $y$, then it serves $(2a-r,2a-s)$ from $2a-x$ with probability $p$ when the taxis are at $2a-x$ and $2a-y$. 

Intuitively, this restriction should be lossless: the line has no
distinguished origin or orientation, and reflecting both an algorithm
and a request sequence preserves all costs, so an
absolute left-right preference should not improve the worst-case ratio.

\begin{proof}[Proof of \cref{thm:memoryless-line-lower-bound} for reflection-invariant algorithms]
    Suppose that $\ALG$ is lazy, memoryless, and reflection-invariant. We may assume that its serving choice ignores the destination $s$ by replacing a request $(r, s)$ with two requests $(r, r), (r,s)$.

Start at configuration $\{0,2\}$ and issue the requests $(1,4),(2,8),(4,16),\dots$; more precisely, for $t = 1,2,\ldots$, request $t$ is $(2^{t-1}, 2^{t+1})$. See~\cref{fig:reflection-invariant-step} for an illustration.

\input{figures/reflection-invariant-step}

On the first request, the offline schedule uses its taxi at $0$, pays $1$, and reaches $\{2,4\}$. It serves every later request for free and finishes request $t$ at $\{2^{t},2^{t+1}\}$.

If the online configuration before request $t$ is $\{0,2^{t}\}$, its serving taxi travels $2^{t-1}$ to the midpoint $2^{t-1}$. Reflection about $2^{t-1}$ gives probability $1/2$ to each taxi. Choosing the taxi at $0$ makes the configurations agree after the request. Choosing the taxi at $2^{t}$ leaves the online configuration at $\{0,2^{t+1}\}$. Once the configurations agree after a request, every subsequent request is free for the algorithm as well.

It follows by induction that the online configuration before request $t$ is $\{0,2^{t}\}$ with probability $2^{1-t}$. Each request therefore contributes $2^{1-t}\cdot2^{t-1}=1$ to the expected online cost. After $m$ requests, the algorithm has expected cost $m$, while the offline schedule has paid only $1$. Thus the competitive ratio is infinite.
\end{proof}

\subsection*{AI Methodology} The authors first noticed, without AI assistance, that memory is necessary to achieve a competitive algorithm. Subsequently, the authors explored various algorithmic and analytic ideas, including ideas involving the maintenance of rankings between the taxis. The algorithm and proof in this paper then arose during communications with ChatGPT-6 Astra. ChatGPT-6.1 Sol and Claude Opus 5.5 were also used for polishing the exposition and creating the figures. The authors take full responsibility for the originality and correctness of all content.

\appendix
\section{Lower Bound for \texorpdfstring{$2$}{2}-Taxi against Adaptive Online Adversaries}
\label{app:k=2}

\cite{CK19} proved that every deterministic algorithm for $2$-taxi has competitive ratio at least~$9$, by a reduction from width-$2$ layered graph traversal. We prove the same bound for randomized algorithms against adaptive online adversaries by embedding the two-branch search game of~\cite[Section~2]{PY91} directly into $2$-taxi on the integer line.

The requests make the pickup location trace a search on the two sides of the origin: the algorithm either advances outward on its current side or pays to cross to the other side. The comparison schedule commits to one side in advance and pays only for progress on that side. Writing $a_i$ for the distance from the origin immediately before crossing $i$, and setting $a_0=0$, the construction will give total costs through crossing $n$ satisfying
\[
\ALG_n=2\sum_{i=1}^n a_i+a_{n-1},\qquad \ADV_n\le a_{n-1}+2.
\]
The algorithm accumulates the lengths of all its excursions, whereas the comparison cost depends only on its progress on one side. We derive these relations and then show that they are incompatible with a competitive ratio below $9$. Write $\operatorname{sgn}(x)\in\{-1,1\}$ for the sign of $x$.

\begin{theorem}\label{thm:two-taxi-lower-bound}
Every randomized algorithm for the hard $2$-taxi problem has competitive ratio at least $9$ against adaptive online adversaries, even on the integer line.
\end{theorem}

\begin{proof}
Suppose an algorithm is $\rho$-competitive for some $1<\rho<9$, with additive constant $\beta$. We may assume that it is lazy and uses a taxi already at $r$ whenever possible, since deferring the movement of a taxi until it is used for a request can only decrease the cost. More formally, a lazy copy can simulate a virtual copy of any algorithm, maintaining a minimum-cost matching between their configurations. When the virtual serving taxi reaches $r$, move its matched real taxi to $r$; if a real taxi is already there, an optimal matching can pair the two taxis at $r$, giving a free pickup. Virtual paid movements increase the matching cost by at most their cost, real pickups decrease it by at least their cost, and simultaneous relocation from $r$ to $s$ leaves it unchanged. Starting from identical configurations, the lazy copy therefore pays no more in total. With deterministic matching tie-breaking, its past is determined by the virtual movement history, so an adaptive-online adversary against the lazy copy is also legal against the virtual algorithm.

\paragraph{Simulating the Two-Branch Search.}
Start both the online algorithm and the adversary at $\{-1,1\}$, issue $(-1,-2)$, and set $r=-1$. Repeatedly issue a pickup test $(r,r)$. After observing that the algorithm used the taxi previously at $x$, issue $(r,x+\operatorname{sgn}(x))$ and set $r=x$. This relocation is a separate request: its destination can depend on the preceding test's outcome, and the taxi now at $r$ serves it for free. It ends one unit beyond the serving taxi's location before the test.

Before each test, one online taxi is at $r+\operatorname{sgn}(r)$, the next integer outward from the current search position. The other is at the next unexplored integer on the opposite side: when the walk leaves a side, its taxi remains one unit beyond the departure point. Choosing the taxi on the current side advances $r$ by one unit; choosing the opposite taxi makes $r$ cross the origin. In either case, the test costs exactly the distance traveled by $r$, and the relocation restores this invariant. Thus each return to a side resumes the outward search at the next unexplored integer.

For each $n\ge1$, stop after the $n$th crossing, including its relocation. Let $a_i\ge1$ be $|r|$ immediately before crossing $i$, and put $a_0=0$. The walk starts at $-1$ and turns at alternating locations of absolute values $a_1,\ldots,a_n$. The initial outward movement costs $a_1-1$, each stretch between successive turning points costs $a_i+a_{i+1}$, and the final crossing ends at distance $a_{n-1}+1$ from the origin, costing $a_n+a_{n-1}+1$. Summing gives
\begin{equation}\label{eq:two-taxi-alg-cost}
\ALG_n=2\sum_{i=1}^n a_i+a_{n-1}.
\end{equation}
This is the cost expression in the classical two-branch sequence game~\cite[Lemmas~2.2--2.3]{PY91}.

\input{figures/two-taxi-lower-bound}

\paragraph{The Adversary Strategy.}
The adversary decides how to serve each request before seeing the algorithm's response. Fix $n$ and commit in advance to the destination side of crossing $n$: right for odd $n$, left for even $n$. On the initial request $(-1,-2)$, the adversary uses the taxi at $-1$ if committed to the left, and the taxi at $1$ otherwise, paying at most $2$. Thereafter, on a pickup test $(r,r)$, it uses the taxi at $r+\operatorname{sgn}(r)$ if $r$ is on its committed side, paying $1$; otherwise, it uses a taxi already at $r$, paying zero. It serves each subsequent relocation request with the taxi at $r$.

These rules maintain the configurations in \cref{fig:two-taxi-configurations}. Before a test, the configurations agree when $r$ is on the committed side; otherwise, the adversary is at $\{r,r+\operatorname{sgn}(r)\}$. Either possible algorithmic choice preserves this invariant after the following relocation. In particular, the taxi prescribed for each test is available before the algorithm responds. After a test, both players have a taxi at $r$, so the prescribed relocation is also available and free. The relocation destination uses only the previous test's outcome, and every adversary serving decision is fixed before the algorithm responds to that request.

The adversary pays once at each integer pickup location on its committed side. Successive visits resume one unit beyond the previous departure point, so these payments cover distinct points. Through crossing $n$, the paid locations have absolute values $1,\ldots,a_{n-1}$. For $n\ge2$, the walk leaves its committed side at crossing $n-1$ and pays nothing until crossing $n$; for $n=1$, there are no unit payments and $a_0=0$. Including the initial cost gives
\begin{equation}\label{eq:two-taxi-adv-cost}
\ADV_n\le a_{n-1}+2.
\end{equation}
The request-generation rule does not use $n$; only the comparison schedule and stopping rule do. Since the algorithm is not told either choice, all stopping indices can be coupled to the same random sequence $(a_i)$.

\paragraph{Stopping at a Crossing.}
We must justify that the prescribed crossing occurs and that its costs have finite expectations. Truncate play after crossing $n$ or after $T$ test--relocation pairs, whichever comes first. Inductively assume $\E[a_{n-1}]<\infty$, starting with $a_0=0$; use the common continuation above to define $a_{n-1}$ even if a truncation stops earlier. The same unit-payment count bounds the comparison cost of every truncation by $a_{n-1}+2$. Competitiveness therefore bounds the expected online cost by $\rho(\E[a_{n-1}]+2)+\beta$, uniformly in $T$.

Every pair costs at least one. If crossing $n$ never occurred, the online cost would consequently tend to infinity as $T\to\infty$. Monotone convergence and the uniform expectation bound rule this out and give finite expected cost through crossing $n$. Equation~\eqref{eq:two-taxi-alg-cost} then gives $\E[a_n]<\infty$, completing the induction and justifying expectations in \eqref{eq:two-taxi-alg-cost}--\eqref{eq:two-taxi-adv-cost}.

\paragraph{Applying the Classical Sequence Argument.}
It remains to show that the two cost bounds cannot hold for $\rho<9$. Competitiveness and \eqref{eq:two-taxi-alg-cost}--\eqref{eq:two-taxi-adv-cost} give
\[
\sum_{i=1}^n \E[a_i]\le q\E[a_{n-1}]+c,
\qquad q=\frac{\rho-1}{2}<4,
\qquad c=\rho+\frac\beta2.
\]
Thus every partial sum of the expected crossing distances is bounded by a multiple of its penultimate term, apart from the fixed constant $c$. To absorb this constant, let $s_n=\sum_{i=1}^n \E[a_i]-c$ for $n\ge0$. Since $a_i\ge1$, these shifted partial sums are eventually positive and strictly increasing, while $s_n\le q(s_{n-1}-s_{n-2})$ for $n\ge2$. For all sufficiently large $n$,
\[
1<\frac{s_n}{s_{n-1}}
\le q\left(1-\frac{s_{n-2}}{s_{n-1}}\right)
\le\frac q4\frac{s_{n-1}}{s_{n-2}},
\]
where the last inequality follows from $(s_{n-1}-2s_{n-2})^2\ge0$. These successive growth ratios must exceed $1$, but the displayed inequality contracts them by a factor of at most $q/4<1$ at every step. Iteration therefore forces a ratio below $1$, a contradiction.
\end{proof}

\section{A Matching Lower Bound for Our Algorithm}\label{sec:lb-algorithm}
Here, we show that the factor $9^{k-1}$ is tight for our algorithm, already on the real line. The construction maintains configurations that differ in only one taxi: the algorithm has an extra taxi at $a$, and the adversary has an extra taxi at $b>a$. The remaining $k-1$ taxis of each player can be paired at identical locations; we call these taxis \emph{common}. By moving common taxis for free in both configurations, the adversary repeatedly arranges requests that can increase the gap $b-a$.

\begin{theorem}\label{thm:lb-algorithm}
For every $k\ge2$, our algorithm has competitive ratio at least $9^{k-1}$ against adaptive online adversaries on the real line.
\end{theorem}

The basic step uses the first $j$ positions of the online ordered list, for some $2\le j\le k$. Arrange $j-1$ common taxis at $c=2b-a$ and the extra online taxi at $a$ in position $j$, then request $(b,b)$. All first $j$ taxis are at distance $b-a$ from $b$. Since $w_j=2\sum_{i<j}w_i$, their local winner is the taxi at $a$ with probability $2/3$ and a common taxi at $c$ with probability $1/3$. If the algorithm serves from $a$, the configurations agree. If it serves from $c$, the taxi arriving at $b$ becomes common, leaving an extra online taxi at $a$ and an extra adversary taxi at $c$. The gap has doubled, so we can repeat the same step with $c$ as the new $b$.

When there are later taxis, we place them much farther away. The global winner then comes from the common taxis at $c$ with probability approaching $1/3$. Thus repeated choices from $c$ double the gap while their probabilities decrease by a factor approaching $1/3$. A later taxi is selected with very small probability, but selecting it starts a continuation whose expected cost is proportional to its distance. These rare choices therefore remain significant in expectation and are essential to obtaining the factor $9^{k-1}$. We now describe a finite construction implementing this idea, then calculate its expected cost.

\begin{proof}[Proof of \cref{thm:lb-algorithm}]
We construct a finite adaptive online adversary that pays exactly one, while the expected online cost can be made arbitrarily close to $9^{k-1}$. Fix a positive integer $N$: for each fixed prefix size $j$, we will issue at most $N$ requests of the form $(b,b)$, rearranging the common taxis before each one. Choose $L>2^{N+3}$ and set $a = -1$, $z_i = L^{i-1}$ for $2 \le i \le k$. Successive locations satisfy $z_{i+1}=Lz_i$. This large spacing will keep the later taxis far from the pickup even after the gap doubles $N$ times.

\paragraph{Starting the Construction.}
Start all taxis at zero. We first prepare identical configurations with online ordered list $(a,z_2,\ldots,z_k)$ at no cost to either player. Recall that, among taxis already at the pickup, the algorithm uses the one appearing first in its ordered list before the update. For $m=k,k-1,\ldots,2$, issue $(0,z_m)$ exactly $m$ times, followed by $(z_m,0)$ exactly $m-1$ times. At the start of this step, the first $m$ taxis are at zero and every later position is already at its prescribed location. The first group moves these $m$ taxis to $z_m$ for free. By the zero-distance tie rule, the second group returns the first $m-1$ positions to zero, leaving position $m$ at $z_m$. Finally issue $(0,a)$ to move the first taxi to $a$. The adversary makes the same free relocations throughout.

Now issue $(0,0)$ and let the adversary use its taxi at $a$, paying one. If the algorithm also serves from $a$, terminate. Otherwise, suppose it serves from $z_j$. The taxi arriving at zero is common, and the configurations now have an extra online taxi at $a$ and an extra adversary taxi at $b=z_j$. Only the first $j$ online positions have changed: they contain the taxi at $a$ and $j-1$ common taxis, while every later position $i>j$ is still at $z_i$ and is common. We call the continuation from this configuration a \emph{level-$j$ stage}. Throughout the construction, a taxi's position means its index in the online ordered list.

\paragraph{Continuing a Stage.}
A level-$j$ stage starts with the configuration just described. During the stage, the extra online location $a$ and all positions $i>j$ remain fixed, while the extra adversary location $b$ may change. We repeat the following preparation and request $(b,b)$ at most $N$ times.

Before issuing $(b,b)$, we arrange the first $j-1$ online positions at $c=2b-a$ and position $j$ at $a$, as in the basic step. First gather the $j-1$ common prefix taxis at $a$. For each such taxi not already at $a$, issue $(p,a)$ at its current location $p$, and let the adversary use its common taxi at $p$. Both players serve these requests for free. The adversary can identify the required locations and multiplicities from the online configuration: remove the extra copy at $a$ and the suffix copies at $z_i$, $i>j$; the remaining copies are precisely the common prefix taxis. Thus this preparation does not require observing the hidden ordered list.

All first $j$ online taxis are now at $a$, and the adversary has $j-1$ common taxis there. Issue $(a,c)$ exactly $j-1$ times, using an adversary taxi already at $a$ each time. After $t$ of these requests, the zero-distance tie rule places the first $t$ online positions at $c$ and leaves positions $t+1,\ldots,j$ at $a$. Hence the required arrangement is obtained for free, regardless of the order of the prefix on entry. Every request in this preparation is served by a prefix taxi already at the pickup, so all suffix positions are preserved.

Issue $(b,b)$, which the adversary serves for free with its extra taxi at $b$. There are three possible outcomes:
\begin{itemize}[itemsep=0em]
 \item If the algorithm serves from $a$, the configurations agree; terminate.
 \item If it serves from $c$, the taxi arriving at $b$ becomes common, while the adversary has an extra taxi at $c$. Set $b\gets c$. The first $j$ positions still contain the extra online taxi at $a$ and $j-1$ common taxis, so we can repeat the preparation with this new $b$. The gap $b-a$ has doubled. Terminate if this was the stage's $N$th request $(b,b)$.
 \item If it serves from $z_i$, $i>j$, the taxi arriving at $b$ becomes common, while the adversary has an extra taxi at $z_i$. Only the first $i$ positions change. They contain the extra online taxi at $a$ and $i-1$ common taxis, and every later position remains at its prescribed location. Thus we have exactly the entry configuration for a level-$i$ stage; start that stage, with a fresh allowance of $N$ requests.
\end{itemize}
In particular, every new stage starts with $b=z_i$, as required. Within a level-$j$ stage, after $n<N$ choices from $c$, the gap is $b-a=2^n(z_j-a)$. For $j<k$, the next common location therefore satisfies $c<2^{N+1}z_j<z_{j+1}$. This is why we chose the spacing $L$ large compared with $2^N$: the common taxis never reach the later fixed locations during the stage. The serving locations $a$, $c$, and $z_i$, $i>j$, are distinct, so observing the serving location identifies which outcome occurred.

Every request after the initial $(0,0)$ costs the adversary zero. Each request and the adversary's serving taxi are chosen before the algorithm responds; any relocation depending on a response is a separate subsequent request. The construction uses only visible locations and multiplicities. A new stage always has a strictly larger level, so at most $k-1$ stages occur, each with at most $N$ requests $(b,b)$ and a bounded number of free preparation requests. Thus, for each finite $N,L$, the entire construction is a finite adaptive online adversary with total cost exactly one.

\paragraph{Expected Cost Starting from a Stage.}
Keep $N$ fixed and let $L\to\infty$; all $o(1)$ terms below refer to this limit. Working backwards through $j=k,k-1,\ldots,2$, we obtain coefficients $A_j$ depending only on $N,k$. We show that a level-$j$ stage, including all subsequent stages, has expected remaining online cost at least $(A_j-o(1))(z_j-a)$. At level $j$, assume these bounds are already known for $i>j$.

Consider the request after $n$ choices from $c$ in the current stage, where $0\le n<N$. Write $D=2^n(z_j-a)$ for the current gap, so $b=a+D$. Immediately before this request, the free preparation puts the first $j-1$ positions at $c$ and position $j$ at $a$, all at distance $D$ from $b$. Since $\sum_{i\le j}w_i=3^{j-1}$, the total exponential rate is
\[
 Z=\frac{3^{j-1}}D+\sum_{i>j}\frac{w_i}{z_i-b}\sim\frac{3^{j-1}}D.
\]
The asymptotic relation holds because $b$ and $D$ are at most a fixed multiple of $z_j$, whereas $z_i/z_j\to\infty$ for every $i>j$. The probability of serving from $c$ therefore tends to $1/3$.

The expected online cost of this one request is exactly $3^{k-1}/Z$. Indeed, a taxi at distance $d$ with weight $w_i$ is chosen with probability $w_i/(dZ)$, so its contribution is $w_i/Z$; summing over all positions gives $\sum_{i=1}^k w_i/Z=3^{k-1}/Z$. Thus even the rare choices of later taxis contribute to the immediate expected cost.

We must also count the remaining cost when a later taxi is chosen. For $i>j$, serving from $z_i$ has probability $w_i/[Z(z_i-b)]$ and starts a level-$i$ stage. Its expected contribution, using the inductive bound, is at least
\[
 \frac{w_i}{Z(z_i-b)}(A_i-o(1))(z_i-a)=\frac{D}{3^{j-1}}\bigl(w_iA_i-o(1)\bigr).
\]
Here $(z_i-a)/(z_i-b)\to1$: the large continuation cost cancels the small selection probability. Adding these contributions to the immediate request cost gives the lower bound
\[
 \frac{D}{3^{j-1}}\left(3^{k-1}+\sum_{i>j}w_iA_i-o(1)\right).
\]
This accounts for the current request and all later stages started by this request. It does not yet count further requests at the current level.

To reach this request, the preceding $n$ requests at level $j$ must all have selected a common taxi at $c$. Conditional on reaching each preceding request, the algorithm chooses from $c$ with probability tending to $1/3$. Hence the probability of reaching request $n+1$ tends to $3^{-n}$. Its gap is $2^n(z_j-a)$. Consequently, its expected contribution has a factor $(2/3)^n$ after normalization by the entry gap $z_j-a$. Summing over $n=0,\ldots,N-1$ counts all requests in the stage and all subsequent stages exactly once. Since
\[
 \sum_{n=0}^{N-1}(2/3)^n=3\bigl(1-(2/3)^N\bigr),
\]
we obtain the desired bound by defining $q\coloneqq1-(2/3)^N$ and
\begin{equation}\label{eq:lb-recurrence}
 A_j=\frac{q}{3^{j-2}}\left(3^{k-1}+\sum_{i=j+1}^k w_iA_i\right),\qquad 2\le j\le k.
\end{equation}
For $j=k$, the sum is empty and this gives $A_k=3q$, starting the induction. The same calculation then gives each preceding coefficient. The estimates are uniform over the finitely many requests and levels, and the free preparation fixes the ordered configuration before each request regardless of the prefix order on entry.

\paragraph{The Resulting Lower Bound.}
For $2\le j<k$, compare \eqref{eq:lb-recurrence} at $j$ and $j+1$. Since $w_{j+1}=2\cdot3^{j-1}$, we obtain
\[
 A_j=3A_{j+1}+\frac{qw_{j+1}}{3^{j-2}}A_{j+1}=3(1+2q)A_{j+1}.
\]
Together with $A_k=3q$, this gives $A_j=3q\,[3(1+2q)]^{k-j}$ for $2\le j\le k$. Thus each additional level contributes a factor $3(1+2q)$, approaching $9$ as $N$ grows.

It remains to count the initial request $(0,0)$ and the stage it starts. Its total rate is $1+\sum_{i=2}^k w_i/z_i$, which tends to one, so its expected immediate cost tends to $3^{k-1}$. Serving from $z_i$ has probability asymptotic to $w_i/z_i$ and starts a stage of expected remaining cost at least $(A_i-o(1))(z_i-a)$. Its contribution is therefore at least $w_iA_i-o(1)$. Hence
\[
 \liminf_{L\to\infty}\E[\ALG]\ge3^{k-1}+\sum_{i=2}^k w_iA_i=\left(\frac1q+2\right)A_2=[3(1+2q)]^{k-1}.
\]
The middle equality follows from \eqref{eq:lb-recurrence} at $j=2$ and $w_2=2$. The adversary pays exactly one. For fixed finite $N$, taking $L$ sufficiently large makes the expected online cost at least this bound minus any prescribed error. Increasing $N$ then makes the bound approach $9^{k-1}$.

Finally, we choose finite parameters and remove the additive constant. Fix $\rho<9^{k-1}$ and an additive constant $\beta$. Choose finite $N$ and then finite $L$ so that the expected online cost is $\rho'>\rho$. Scale every location by a factor $D>0$. All exponential rates are divided by $D$, preserving the distribution of serving choices and ordered-list updates, while both costs are multiplied by $D$. The initial configuration remains all zero, so the same additive constant applies. Taking $(\rho'-\rho)D>\beta$ gives $\E[\ALG]>\rho\,\E[\ADV]+\beta$.
\end{proof}

\section{The Memoryless Lower Bound Without Reflection Invariance}
\label{app:memoryless}
We prove \cref{thm:memoryless-line-lower-bound} without the reflection invariance assumption used in \cref{sec:memoryless}. The midpoint construction there relies on equally likely serving choices. Without symmetry, those choices may be biased. We instead compare the same online configuration and request with two different offline configurations. Memorylessness forces the same serving probability in both comparisons, giving constraints that no bounded discrepancy cost can satisfy.

\memorylesslowerbound*

Suppose that $\ALG$ is a lazy memoryless algorithm with competitive ratio $1\le \rho<\infty$ and additive constant $\beta$.

\paragraph{An Oriented Discrepancy Cost.}
When the configurations differ in one taxi, write $x$ for the extra online location and $y$ for the extra offline location; the other taxi is common. We will measure the expected online cost that this discrepancy can cause while the offline schedule pays nothing. The common taxi can be relocated to $x$ for free in both configurations, giving online configuration $\{x,x\}$ and offline configuration $\{x,y\}$. A reverse free relocation restores its former location. By memorylessness, these relocations do not affect the supremum of future costs, so we use these configurations to define our discrepancy cost.

Let $\mathrm{free}(\{x,y\})$ denote the set of finite request sequences $\sigma$ that the offline configuration $\{x,y\}$ can serve for free. Define $F\colon\R^2\to[0,\infty]$ by
\[
F(x,y)\coloneqq\sup_{\sigma \in \mathrm{free}(\{x, y\})}\E[\ALG(\sigma)],
\]
where $\ALG$ starts at $\{x,x\}$. The order of the arguments matters, and $F(x,y)$ need not equal $F(y,x)$.

To create this discrepancy from identical configurations, start both players at $\{x,x\}$ and issue $(x,y)$ followed by $(y,x)$. The starting configuration can itself be obtained for free from any common initial configuration by requesting each of the two initial taxi locations with destination $x$. After $(x,y)$, both configurations are $\{x,y\}$. By laziness, the algorithm then uses its taxi at $y$ and returns to $\{x,x\}$ for free. The offline schedule instead uses its taxi at $x$, pays $|x-y|$, and finishes at $\{x,y\}$. Repeating this preparation and a suitable free offline continuation will bound $F$ by $\rho$ times the preparation cost.

\begin{lemma}\label{lem:memoryless-bounds}
The function $F$ satisfies
\[
|x-y|\le F(x,y)\le \rho|x-y|
\qquad(x,y\in\mathbb R).
\]
\end{lemma}
\begin{proof}
The request $(y,y)$ costs $|x-y|$ online and zero offline, proving the lower bound.

For the upper bound, fix a sequence $\sigma$ in the definition of $F(x,y)$. While the offline schedule serves it for free, the configurations either coincide or their only unmatched online taxi remains at $x$. Indeed, when the algorithm has a taxi at the pickup, both players make the same free relocation. Otherwise, the algorithm either uses its extra taxi at $x$ and makes the configurations agree, or uses the common taxi and leaves the unmatched online taxi at $x$.

If the final offline configuration is $\{u,v\}$, append $(u,x),(v,x)$. These requests return the offline configuration to $\{x,x\}$ for free. The same discrepancy invariant shows that the online configuration also returns to $\{x,x\}$; its reset may have positive cost, which only strengthens the lower bound on its total cost.

Repeat the preparation, $\sigma$, and the reset $n$ times. This is one finite request sequence fixed in advance: the free offline trajectory, including $u$ and $v$, is determined by $\sigma$. Memorylessness makes the expected cost of $\sigma$ the same on every repetition, since its online starting configuration is always $\{x,x\}$. The offline schedule pays $n|x-y|$, so competitiveness gives
\[
n\E[\ALG(\sigma)]\le \rho n|x-y|+\beta.
\]
Letting $n\to\infty$ removes the additive constant. Taking the supremum over $\sigma$ proves the upper bound and, in particular, shows that $F$ is finite.
\end{proof}

The second constraint comes from an online configuration $\{x,z\}$ and request $(y,y)$, for three distinct points $x,y,z$. Let $p(x,y,z)$ be the probability of serving from $x$. With offline configuration $\{y,z\}$, choosing $x$ makes the configurations agree, whereas choosing $z$ leaves an extra online taxi at $x$ and an extra offline taxi at $z$. With offline configuration $\{x,y\}$, choosing $z$ makes the configurations agree, whereas choosing $x$ leaves an extra online taxi at $z$ and an extra offline taxi at $x$. The algorithm sees the same configuration and request in both cases, so the two comparisons use the same $p(x,y,z)$.

\begin{lemma}\label{lem:memoryless-comparisons}
For every three distinct points $x,y,z$, there is a $p=p(x,y,z)\in[0,1]$ such that
\[
\begin{aligned}
F(x,y)&\ge p|x-y|+(1-p)\bigl(|z-y|+F(x,z)\bigr),\\
F(z,y)&\ge(1-p)|z-y|+p\bigl(|x-y|+F(z,x)\bigr).
\end{aligned}
\]
\end{lemma}
\begin{proof}
For the first comparison, precede the request $(y,y)$ by the free request $(x,z)$. Starting from the defining configurations for $F(x,y)$, this gives online configuration $\{x,z\}$ and offline configuration $\{y,z\}$. The two serving choices have the costs and discrepancies described above.

To realize the continuation cost $F(x,z)$, fix a sequence approaching its supremum and append it after the request $(y,x)$. If the discrepancy survives, this free request changes online configuration $\{x,y\}$ and offline configuration $\{z,y\}$ into the defining configurations $\{x,x\}$ and $\{x,z\}$. If the configurations have agreed, the same request instead puts both at $\{x,z\}$, and the chosen continuation is free for both. Thus one continuation, chosen before the serving choice is observed, gives the first inequality in \cref{lem:memoryless-comparisons}. An arbitrarily close approximation to the supremum suffices.

For the second comparison, start from the defining configurations for $F(z,y)$ and use the free request $(z,x)$ to reach online configuration $\{x,z\}$ and offline configuration $\{x,y\}$. If serving $(y,y)$ leaves the discrepancy, append $(y,z)$ to obtain online configuration $\{z,z\}$ and offline configuration $\{z,x\}$, then a fixed continuation approaching $F(z,x)$. On the branch where the configurations agree, the same continuation is free. This proves the second inequality with the same $p(x,y,z)$. Each comparison uses a request sequence fixed in advance; neither continuation is selected after observing the random outcome.
\end{proof}

We now show that the bounds and comparisons in \cref{lem:memoryless-bounds,lem:memoryless-comparisons} are incompatible. The difficulty is that $F$ is oriented: the costs of the two directions of a gap may differ. The next argument uses backward-oriented costs to find short steps, then the two comparisons force the forward-oriented normalized cost to increase.

\begin{lemma}\label{lem:memoryless-incompatibility}
For any finite $\rho\ge1$, no function $F\colon\mathbb R^2\to[0,\infty)$ satisfies the bounds in \cref{lem:memoryless-bounds} and the comparisons in \cref{lem:memoryless-comparisons}.
\end{lemma}
\begin{proof}
We will choose a left endpoint $x$ and move the right endpoint $y$ towards it, tracking $F(x,y)/(y-x)\in[1,\rho]$. An averaging argument finds an interval with few exceptional points. At every other point used by our walk, we can take a short step that increases this ratio by at least the logarithm of the factor by which the step reduces the gap. At exceptional points, we take unit steps; their total shortfall from this logarithmic increase is bounded only in terms of $\rho$. The logarithms telescope to the logarithm of the ratio between the initial and final gaps. Making that ratio sufficiently large forces an increase exceeding $\rho-1$, a contradiction.

\paragraph{Finding the Interval.}
Fix an integer $R>2$, which will be the ratio of spatial scales, and an integer $L\ge \rho R$, providing many scales over which to average. For a fixed right endpoint $y$, we examine the smallest normalized cost of a backward gap as the allowed gap length grows. For each $y\in\{1,\ldots,R^L\}$, define
\[
a_i(y)\coloneqq\min_{t\in\{1,\ldots,R^i\}}\frac{F(y,y-t)}t
\qquad(0\le i\le L).
\]
These minima lie in $[1,\rho]$ and are nonincreasing in $i$, since increasing $i$ allows more candidate lengths. Their total decrease is at most $\rho-1$. For $1\le i\le L$, call $y$ exceptional at scale $i$ if $a_{i-1}(y)>a_i(y)+1$. Each exceptional scale contributes a decrease greater than $1$, so a point is exceptional at fewer than $\rho$ scales.

Averaging over the $L$ scales, some scale $i$ has at most $\rho R^L/L\le R^{L-1}$ exceptional points. Now partition $\{1,\ldots,R^L\}$ into $R^{L-i}$ blocks of $R^i$ consecutive points. Averaging over these blocks gives one with at most $R^{i-1}$ exceptional points. Write it as
\[
\{x+1,\ldots,x+RT\},\qquad T=R^{i-1}.
\]

At a nonexceptional point $y=x+h$ in this block, choose an integer $1\le t\le T$ attaining $a_{i-1}(y)$. The whole backward gap from $y$ to $x$ has length $h\le RT=R^i$, so it is one of the candidates defining $a_i(y)$. Hence
\[
\frac{F(y,y-t)}t=a_{i-1}(y)
\le a_i(y)+1
\le\frac{F(y,x)}h+1.
\]
Thus a short backward gap has normalized cost at most $1$ above that of the whole backward gap. We use this bound on $F(y,y-t)$ to control a serving probability, which in turn forces an increase in the opposite orientation $F(x,y)$.

\paragraph{Shortening the Gap.}
Starting at $h=RT$, decrease $h$ until $h\le2T$. At $y=x+h$, take the step $h\gets h-t$ chosen above if $y$ is nonexceptional, and take the unit step $h\gets h-1$ otherwise. Every step has positive integer length at most $T$, so the walk ends at some $h_*$ with $T<h_*\le2T$. Before every step, $h>2T$, and therefore $t<h/2$ at a nonexceptional point.

For a nonexceptional step, apply \cref{lem:memoryless-comparisons} to $x,y-t,y$, with $p=p(x,y-t,y)$. The second comparison involves the backward-oriented costs and gives
\[
p\bigl(F(y,x)+h-2t\bigr)
\le F(y,y-t)-t
\le\frac thF(y,x).
\]
Since $h\ge2t$, the factor multiplying $p$ is at least $F(y,x)>0$, so $p\le t/h$. The first comparison reads
\[
F(x,y-t)\ge F(x,y)+t-p\bigl(F(x,y)-h+2t\bigr).
\]
Because $F(x,y)\ge h$, its right-hand side is nonincreasing in $p$. Substituting $p\le t/h$ and dividing by $h-t$ yields
\[
\frac{F(x,y-t)}{h-t}-\frac{F(x,y)}h
\ge\frac{2t}h
\ge\ln\frac h{h-t},
\]
where the last inequality uses $t/h\le1/2$. The logarithm records the multiplicative reduction in the gap.

\Needspace{9\baselineskip}
At an exceptional point, we do not have the short-gap bound, so we use a unit step. The second comparison, now with $p=p(x,y-1,y)$, together with $F(y,y-1)\le \rho$ and $F(y,x)\ge h$, gives $p\le(\rho-1)/(2h-2)\le \rho/h$. The first comparison and $F(x,y)\le \rho h$, together with $\ln(h/(h-1))\le1/(h-1)$, give the logarithmic lower bound up to an error:
\[
\frac{F(x,y-1)}{h-1}-\frac{F(x,y)}h
\ge-\frac{\rho^2}{h-1}
\ge\ln\frac h{h-1}-\frac{\rho^2+1}{h-1}.
\]

Since $h-1\ge2T$, the error at each exceptional step is at most $(\rho^2+1)/(2T)$. The walk visits at most $T$ exceptional points because $h$ strictly decreases, so the total error is at most $(\rho^2+1)/2$, independently of $R$. Adding the step bounds makes the logarithms telescope:
\[
\frac{F(x,x+h_*)}{h_*}
-\frac{F(x,x+RT)}{RT}
\ge\ln\frac{RT}{h_*}-\frac{\rho^2+1}{2}
\ge\ln(R/2)-\frac{\rho^2+1}{2}.
\]
The left-hand side is at most $\rho-1$. Choosing $R$ sufficiently large gives a contradiction.
\end{proof}

By \cref{lem:memoryless-bounds,lem:memoryless-comparisons}, the discrepancy cost $F$ satisfies both required properties. Applying \cref{lem:memoryless-incompatibility} therefore contradicts the assumed finite competitive ratio $\rho$, proving \cref{thm:memoryless-line-lower-bound}.

\printbibliography
\end{document}

%% file: preamble.tex
\newtheorem{theorem}{Theorem}[section]
\newtheorem{corollary}[theorem]{Corollary}
\newtheorem{lemma}[theorem]{Lemma}

\newtheorem{claim}[theorem]{Claim}

\crefname{claim}{Claim}{Claims}
\Crefname{claim}{Claim}{Claims}
\crefname{lemma}{Lemma}{Lemmas}
\Crefname{lemma}{Lemma}{Lemmas}

\usepackage{fancyhdr}

\newenvironment{proof-sketch}{\noindent{\bf Sketch of Proof}\hspace*{1em}}{\qed\bigskip}
\newenvironment{proof-idea}{\noindent{\bf Proof Idea}\hspace*{1em}}{\qed\bigskip}
\newenvironment{proof-of-lemma}[1]{\noindent{\bf Proof of Lemma #1}\hspace*{1em}}{\qed\bigskip}
\newenvironment{proof-attempt}{\noindent{\bf Proof Attempt}\hspace*{1em}}{\qed\bigskip}

\makeatletter
\def\fnum@figure{{\bf Figure \thefigure}}
\def\fnum@table{{\bf Table \thetable}}
\long\def\@mycaption#1[#2]#3{\addcontentsline{\csname
  ext@#1\endcsname}{#1}{\protect\numberline{\csname
  the#1\endcsname}{\ignorespaces #2}}\par
  \begingroup
    \@parboxrestore
    \small
    \@makecaption{\csname fnum@#1\endcsname}{\ignorespaces #3}\par
  \endgroup}
\def\mycaption{\refstepcounter\@captype \@dblarg{\@mycaption\@captype}}
\makeatother

\newcommand{\mathify}[1]{\ifmmode{#1}\else\mbox{$#1$}\fi}
\newcommand{\bigO}O
\newcommand{\set}[1]{\mathify{\left\{ #1 \right\}}}

\newcommand{\remove}[1]{}
\newcommand{\ignore}[1]{}

\def\R{\mathbb{R}}

\newcommand{\mc}{\mathcal}

%% file: figures/permutation-update.tex
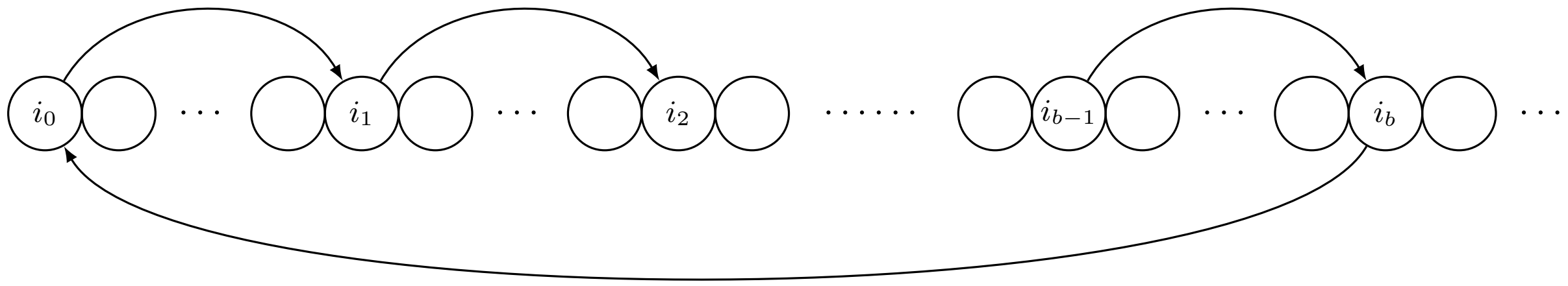
\begin{figure}[t]
\centering
\begin{tikzpicture}[
  x=7.5mm,
  perm slot/.style={circle,draw,line width=.6pt,minimum size=7.5mm,
    inner sep=0pt,font=\footnotesize},
  perm move/.style={-latex,line width=.6pt},
]
  \foreach \x in {1,3.3,5.3,7.6,9.6,12.9,14.9,17.2,19.2}
    \node[perm slot] at (\x,0) {};
  \foreach \x in {2.15,6.45,16.05,20.35}
    \node at (\x,0) {$\cdots$};
  \node at (11.25,0) {$\cdots\cdots$};
  \foreach \x/\name/\lab in
      {0/i0/{i_0},4.3/i1/{i_1},8.6/i2/{i_2},13.9/ib1/{i_{b-1}},18.2/ib/{i_b}}
    \node[perm slot] (\name) at (\x,0) {$\lab$};
  \draw[perm move] (i0)  to[out=60,in=120] (i1);
  \draw[perm move] (i1)  to[out=60,in=120] (i2);
  \draw[perm move] (ib1) to[out=60,in=120] (ib);
  \draw[perm move] (ib) to[out=-120,in=-60,looseness=.4] (i0);
\end{tikzpicture}
\caption{Changes in positions during the update of the ordered list. Here $i_0=1$, $i_{a+1}$ is the first index $i>i_a$ with $\zeta_i<\zeta_{i_a}$, and $x_{i_b}$ is the global winner. The global winner moves to the front, each taxi $x_{i_a}$ with $0\le a<b$ moves back to position $i_{a+1}$, and all other taxis keep their positions.}
\label{fig:permutation-update}
\end{figure}

%% file: figures/exchange-optimal.tex
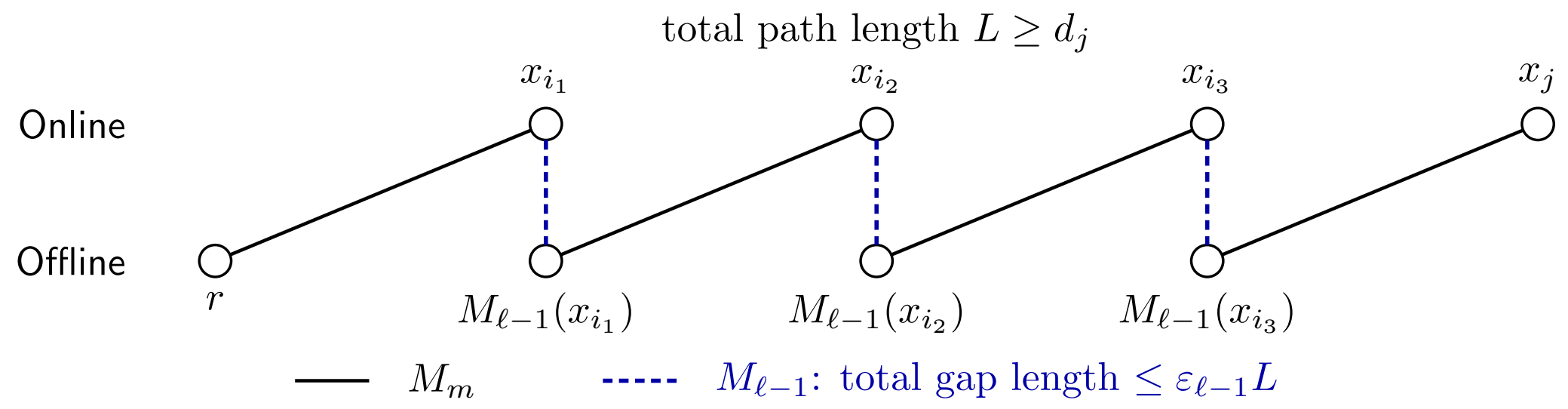
\begin{figure}[t]
\centering
\begin{tikzpicture}[
  font=\small,
  exchange point/.style={
    circle,draw=black,fill=white,
    minimum size=2.8mm,inner sep=0pt,outer sep=0pt,line width=.65pt
  },
  new matching edge/.style={draw=black,line width=.85pt},
  old matching gap/.style={
    draw=blue!65!black,dash pattern=on 2.4pt off 1.6pt,line width=1.05pt
  },
  exchange row label/.style={font=\small\sffamily,anchor=east}
]
  \node[exchange row label] at (-.65,1.2) {Online};
  \node[exchange row label] at (-.65,0) {Offline};
  \node at (5.8,2) {total path length $L\ge d_j$};

  \node[exchange point,label=below:{$r$}] (r) at (0,0) {};
  \node[exchange point,label=above:{$x_{i_1}$}] (a1) at (2.9,1.2) {};
  \node[exchange point,label=above:{$x_{i_2}$}] (a2) at (5.8,1.2) {};
  \node[exchange point,label=above:{$x_{i_3}$}] (a3) at (8.7,1.2) {};
  \node[exchange point,label=above:{$x_j$}] (j) at (11.6,1.2) {};
  \node[exchange point,label=below:{$M_{\ell-1}(x_{i_1})$}] (b1) at (2.9,0) {};
  \node[exchange point,label=below:{$M_{\ell-1}(x_{i_2})$}] (b2) at (5.8,0) {};
  \node[exchange point,label=below:{$M_{\ell-1}(x_{i_3})$}] (b3) at (8.7,0) {};

  \draw[new matching edge] (r)--(a1);
  \draw[new matching edge] (b1)--(a2);
  \draw[new matching edge] (b2)--(a3);
  \draw[new matching edge] (b3)--(j);
  \foreach \h in {1,2,3}
    \draw[old matching gap] (a\h)--(b\h);

  \draw[new matching edge] (.7,-1.05)--(1.35,-1.05);
  \node[anchor=west] at (1.55,-1.05) {$M_m$};
  \draw[old matching gap] (3.4,-1.05)--(4.05,-1.05);
  \node[anchor=west,text=blue!65!black] at (4.25,-1.05)
    {$M_{\ell-1}$: total gap length $\le\varepsilon_{\ell-1}L$};
\end{tikzpicture}
\caption{The alternating path in $M_m\cup M_{\ell-1}$. The dashed old matching edges form the gaps between consecutive $M_m$-edges. Here $i_1,i_2,i_3<\ell\le j\le m$. Their online endpoints are within distance $L$ of $r$, so their
total length is at most $\varepsilon_{\ell-1}L$. Moving $x_j$ to $r$,
matching it there at zero cost, and replacing the solid path edges by the old
edges decreases the level-$m$ matching cost by at least $(1-2\varepsilon_{\ell-1})L$.}
\label{fig:optimal-matching-exchange}
\end{figure}

%% file: figures/epsilon-radius-entry.tex
\begin{figure}[tbp]
\centering
\begin{tikzpicture}[
  font=\small,
  radius point/.style={
    circle,draw=black,fill=white,
    minimum size=2.8mm,inner sep=0pt,outer sep=0pt,line width=.65pt
  },
  radius new edge/.style={-latex,draw=black,line width=.85pt},
  radius old edge/.style={
    -latex,draw=blue!65!black,
    dash pattern=on 2.4pt off 1.6pt,line width=1.05pt
  },
  radius row label/.style={font=\small\sffamily,anchor=east}
]
  \filldraw[fill=black!4,draw=black!25,rounded corners=2pt]
    (6.35,-.65) rectangle (11.75,2.65);
  \node[font=\footnotesize] at (9.05,2.45)
    {distance to $r$ at most $T$};

  \node[radius row label] at (-.7,1.4) {Online};
  \node[radius row label] at (-.7,0) {Offline};

  \node[radius point,label=above:{$x_m$}] (xm) at (0,1.4) {};
  \node[font=\footnotesize] at (0,2.45) {start: $d_m>T$};
  \node[radius point,label=above:{$x_h$}] (xh) at (3.6,1.4) {};
  \node[radius point,label=above:{$x_i$ (first online inside)}] (xi) at (8.2,1.4) {};
  \node[radius point,label=below:{$M_{m-1}(x_h)$}] (yh) at (1.8,0) {};
  \node[radius point,label=below:{$M_{m-1}(x_i)$}] (yi) at (5.3,0) {};
  \node[radius point,label=below:{$r$}] (r) at (10.6,0) {};

  \draw[radius new edge] (xm)--(yh);
  \draw[radius old edge] (yh)--(xh);
  \draw[radius new edge] (xh)--(yi);
  \draw[radius old edge] (yi)--(xi)
    node[midway,right=4pt,text=blue!65!black] {$\le C_{m-1}(T)$};

  \draw[decorate,decoration={brace,amplitude=4pt},line width=.55pt]
    (0,3.15)--(8.2,3.15)
    node[midway,above=6pt] {initial subpath from $x_m$ to $x_i$: length $\ge d_m-T$};

  \draw[draw=black,-latex,line width=.85pt] (1.1,-1.05)--(1.75,-1.05);
  \node[anchor=west] at (1.95,-1.05) {$M_m$};
  \draw[draw=blue!65!black,-latex,dash pattern=on 2.4pt off 1.6pt,line width=1.05pt]
    (4.4,-1.05)--(5.05,-1.05);
  \node[anchor=west,text=blue!65!black] at (5.25,-1.05) {$M_{m-1}$};
\end{tikzpicture}
\caption{An initial subpath of $P$ for $T<d_m$, with $1\le h,i<m$.
It starts at $x_m$ and stops at the first online vertex $x_i$ within distance
$T$ of $r$ (shaded). The new edge ending at $M_{m-1}(x_i)$ is counted in
$L_m^{>T}$ because its online endpoint $x_h$ is outside; the final old edge
is counted by $C_{m-1}(T)$. }
\label{fig:epsilon-radius-entry}
\end{figure}
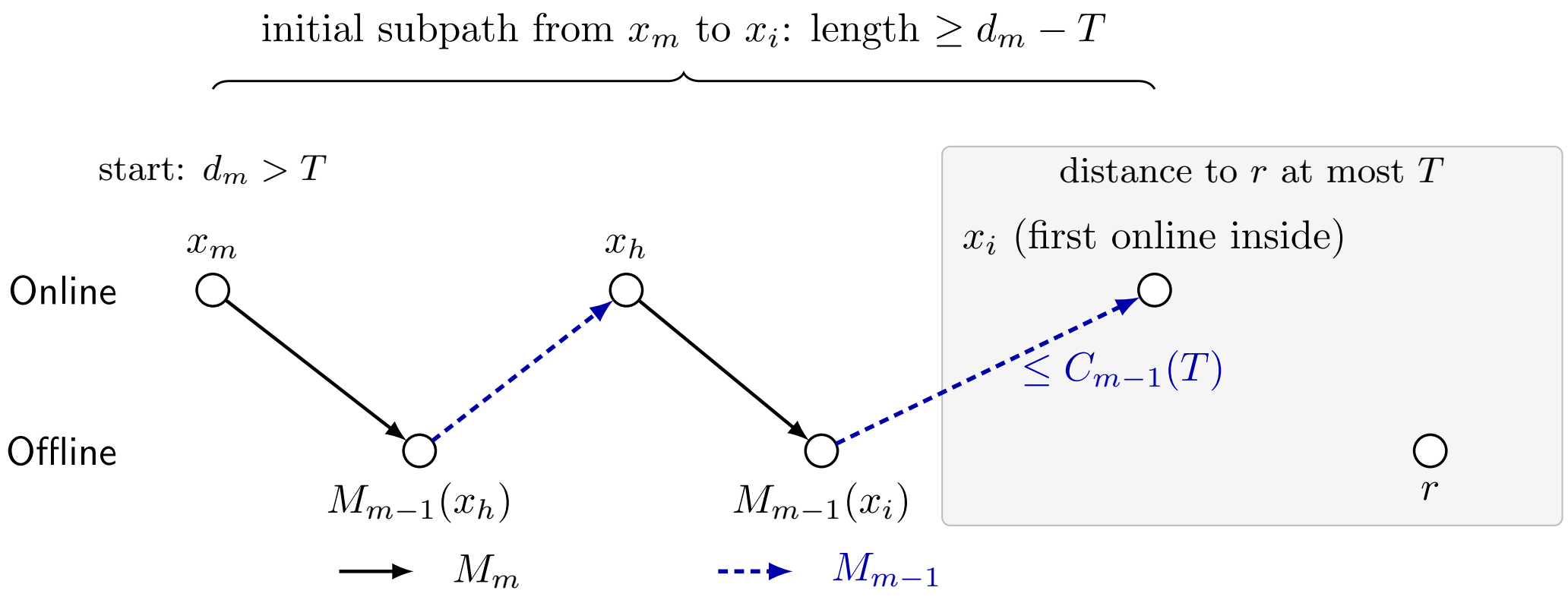

%% file: figures/reflection-invariant-step.tex
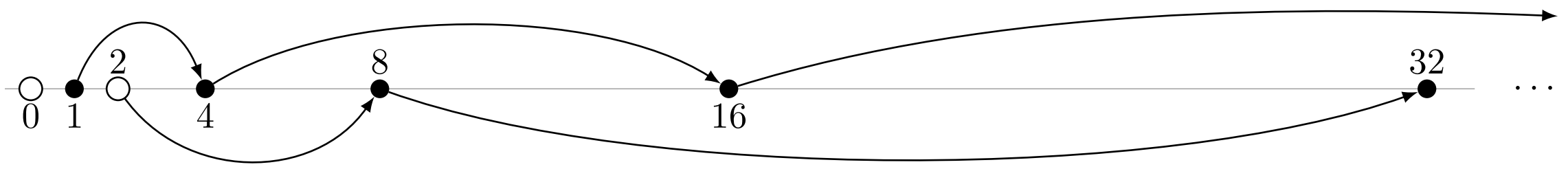
\begin{figure}[ht]
\centering
\begin{tikzpicture}[
  font=\small,
  request point/.style={circle,fill=black,minimum size=1.8mm,
    inner sep=0pt,outer sep=0pt},
  initial taxi/.style={circle,draw=black,fill=white,minimum size=2.2mm,
    inner sep=0pt,outer sep=0pt,line width=.5pt},
  request arrow/.style={-latex,draw=black,line width=.5pt}
]
  \draw[black!35,line width=.3pt] (-.25,0)--(13.9,0);
  \foreach \x in {0,1,2,4,8,16,32}
    \node[request point] (p\x) at ({.42*\x},0) {};
  \foreach \x in {0,1,4,16}
    \node[inner sep=1pt,below=3pt] at (p\x) {$\x$};
  \foreach \x in {2,8,32}
    \node[inner sep=1pt,above=3pt] at (p\x) {$\x$};
  \node at (14.5,0) {$\cdots$};

  \draw[request arrow] (p1) .. controls (.735,.8) and (1.365,.8) .. (p4);
  \draw[request arrow] (p2) .. controls (1.47,-.9) and (2.73,-.9) .. (p8);
  \draw[request arrow] (p4) .. controls (2.94,.8) and (5.46,.8) .. (p16);
  \draw[request arrow] (p8) .. controls (5.88,-.9) and (10.92,-.9) .. (p32);
  \draw[request arrow] (p16) .. controls (9.3,.8) and (12.2,.8) .. (14.7,.7);

  \foreach \x in {0,2}
    \node[initial taxi] at (p\x) {};
\end{tikzpicture}
\caption{Illustration of the requests. The white circle marks initial taxis. The adversary serves the first request using the taxi at $0$ and continues for free. Conditional on being at $\{0, 2^t\}$ before request $t$, the online algorithm faces a symmetric pickup and uses the ``wrong'' taxi on the right with probability $1/2$.}
\label{fig:reflection-invariant-step}
\end{figure}

%% file: figures/two-taxi-lower-bound.tex
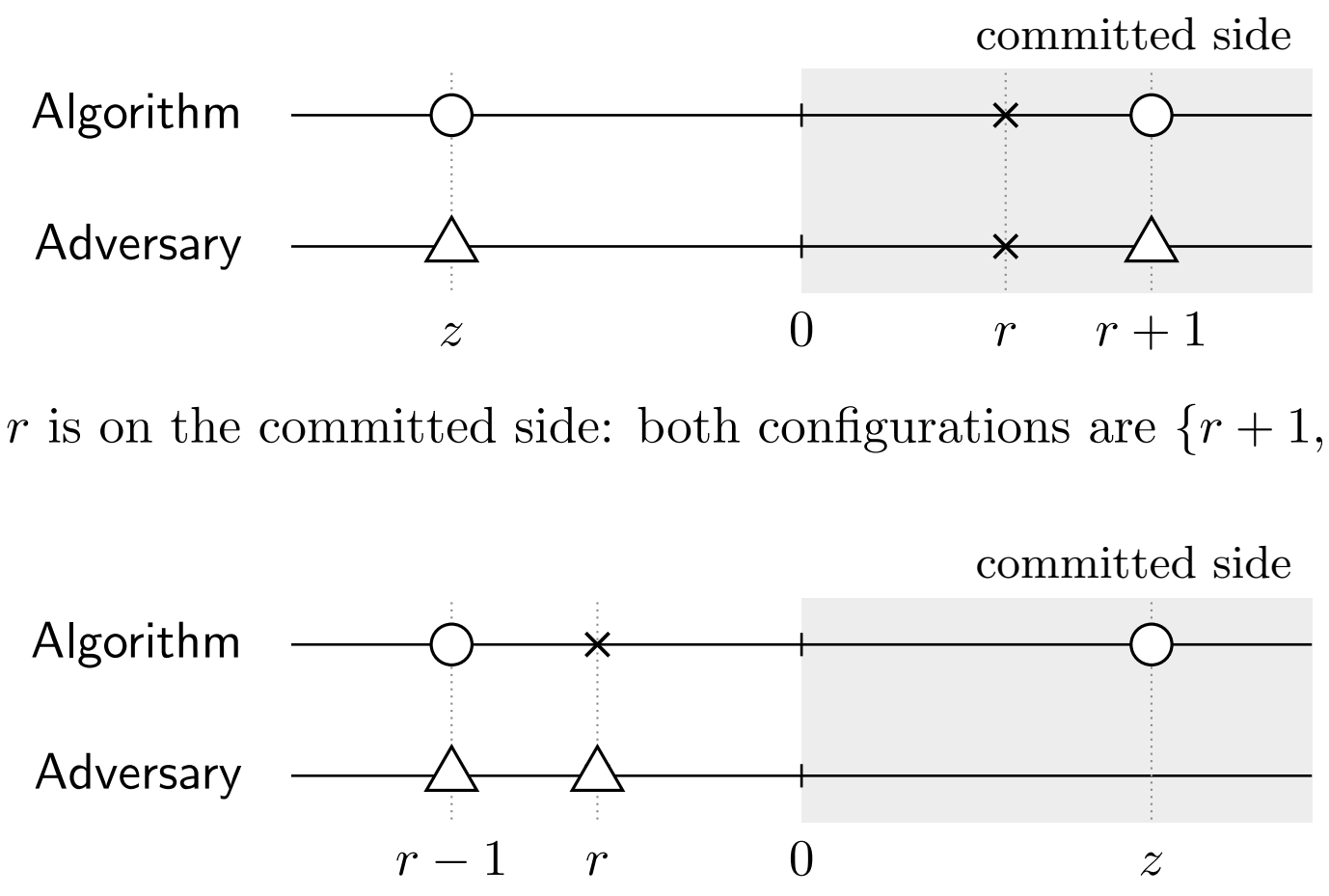
\begin{figure}[t]
\centering
\tikzset{
  two taxi line/.style={line width=.5pt},
  two taxi guide/.style={densely dotted,line width=.4pt,black!45},
  two taxi alg/.style={circle,draw,fill=white,line width=.6pt,inner sep=0pt,minimum size=2.8mm},
  two taxi adv/.pic={\filldraw[fill=white,line width=.6pt]
    (90:2mm) -- (210:2mm) -- (330:2mm) -- cycle;},
  two taxi request/.pic={\draw[line width=.7pt]
    (-.08,-.08) -- (.08,.08) (-.08,.08) -- (.08,-.08);},
  two taxi axes/.pic={
    \fill[black!7] (0,-.32) rectangle (3.5,1.22);
    \node[font=\footnotesize,anchor=south east] at (3.5,1.22) {committed side};
    \foreach \h/\who in {.9/Algorithm,0/Adversary}{
      \draw[two taxi line] (-3.5,\h) -- (3.5,\h);
      \draw[two taxi line] (0,\h-.08) -- (0,\h+.08);
      \node[font=\small\sffamily,anchor=east] at (-3.7,\h) {\who};
    }
  },
  two taxi pos/.style={font=\small,anchor=base},
}
\begin{subfigure}{\textwidth}
\centering
\begin{tikzpicture}
  \pic {two taxi axes};
  \foreach \x in {-2.4,1.4,2.4} \draw[two taxi guide] (\x,-.3) -- (\x,1.2);
  \pic at (1.4,.9) {two taxi request};
  \pic at (1.4,0) {two taxi request};
  \node[two taxi alg] at (-2.4,.9) {};
  \node[two taxi alg] at (2.4,.9) {};
  \pic at (-2.4,0) {two taxi adv};
  \pic at (2.4,0) {two taxi adv};
  \foreach \x/\lab in {-2.4/{$z$},0/{$0$},1.4/{$r$},2.4/{$r+1$}}
    \node[two taxi pos] at (\x,-.68) {\lab};
\end{tikzpicture}
\caption{$r$ is on the committed side: both configurations are $\{r+1,z\}$.}
\label{fig:two-taxi-committed}
\end{subfigure}

\bigskip

\begin{subfigure}{\textwidth}
\centering
\begin{tikzpicture}
  \pic {two taxi axes};
  \foreach \x in {-2.4,-1.4,2.4} \draw[two taxi guide] (\x,-.3) -- (\x,1.2);
  \pic at (-1.4,.9) {two taxi request};
  \pic at (-1.4,0) {two taxi request};
  \node[two taxi alg] at (-2.4,.9) {};
  \node[two taxi alg] at (2.4,.9) {};
  \pic at (-2.4,0) {two taxi adv};
  \pic at (-1.4,0) {two taxi adv};
  \foreach \x/\lab in {-2.4/{$r-1$},-1.4/{$r$},0/{$0$},2.4/{$z$}}
    \node[two taxi pos] at (\x,-.68) {\lab};
\end{tikzpicture}
\caption{$r$ is on the uncommitted side: the algorithm is at $\{r-1,z\}$, and the adversary is at $\{r,r-1\}$.}
\label{fig:two-taxi-uncommitted}
\end{subfigure}
\caption{The adversary configurations before $(r,r)$, with the committed side drawn as $\mathbb Z_{>0}$ and $z$ denoting the online taxi on the opposite side of the origin from $r$. Circles mark the algorithm's taxis, triangles mark the adversary's taxis, and crosses mark the pickup location $r$. The adversary serves $(r,r)$ at cost $1$ in~(a), with its taxi at $r+1$, and at cost $0$ in~(b), with its taxi at $r$. Not drawn to scale.}
\label{fig:two-taxi-configurations}
\end{figure}